%% file: main.tex
\documentclass{llncs}
\usepackage[T1]{fontenc}
\usepackage{microtype}
\usepackage{amsmath,amssymb,booktabs,array,needspace,placeins}
\usepackage{tikz}
\usepackage[hidelinks,pdfusetitle]{hyperref}
\newcolumntype{P}[1]{>{\raggedright\arraybackslash}p{#1}}
\newcommand{\F}{\mathbb F}
\newcommand{\C}{\mathcal C}
\newcommand{\dist}{\operatorname{dist}}
\newcommand{\wt}{\operatorname{wt}}
\newcommand{\rank}{\operatorname{rank}}
\newcommand{\eps}{\epsilon}
\newcommand{\Ext}{\mathsf{Ext}}
\title{Fast-Encodable Locally Testable Codes for Collective Data Availability Sampling}
\author{Zihao Zhang, Kexi Huang, Jiaheng Zhang, Sihuang Hu}
\institute{}
\newcommand{\Badval}{\mathsf{Bad}_{\rm val}}
\newcommand{\kinit}{k_{\rm init}}
\newcommand{\ill}{\mathsf{ill}}
\newcommand{\inc}{\mathsf{inc}}
\newcommand{\Qhash}{\mathsf{Q}_{\rm hash}}

\begin{document}
\raggedbottom
\maketitle
\thispagestyle{plain}
\begin{abstract}
In collective data availability sampling (DAS), clients check block availability with small samples and pool their observations for recovery. Erasure code commitments ensure that the observed symbols agree with a codeword, as erasure decoding requires. A plain Merkle commitment can authenticate symbols from an invalid encoding, leaving recovery with errors as well as erasures.

We show how to use a single Merkle commitment without a trusted setup by reusing local parity equations for both testing and recovery. The equations define a locally testable code and are imposed on overlapping constant-size sets of coordinates called lines. Each client samples one line, authenticates its symbols, and checks its parity equations. The extractor pools the authenticated observations and tries to recover a codeword using the same equations. It returns the corresponding message only if it recovers a codeword and the numbers of unobserved coordinates and of disagreements with the observations lie within their prescribed budgets.

We prove completeness, soundness, and consistency in the classical random-oracle model for sufficiently large code length $N$. At security parameter $\lambda$, it suffices to use $O(N+\lambda)$ clients, each drawing a uniform line independently after the commitment is fixed. With adaptive publisher replies, the probability that all clients accept and extraction fails is at most $2^{-\lambda}$. When the experiment's total oracle-query budget is $2^{O(\lambda)}$, the commitment uses $O(\lambda)$ bits, and a line request and its authenticated reply together use $O(\lambda\log N)$ bits.

Our code families achieve arbitrarily high constant rates and positive relative distance over any fixed finite field. After a reusable public preprocessing step, a systematic encoder maps messages bijectively to codewords using $O(N\log N)$ field and index operations. The algebraic recovery algorithm uses $O(N)$ such operations on the pooled observations.
\end{abstract}

\input{sections/01-introduction/introduction}
\input{sections/02-overview/overview}
\input{sections/03-model/model}
\input{sections/03-model/generic-security}

\input{sections/04-code/ltc}
\input{sections/04-code/completion}
\input{sections/05-recovery/recovery}
\input{sections/06-das/security}

\FloatBarrier
\label{main-text-end}
\clearpage
\label{references-start}
\bibliographystyle{splncs04}
\bibliography{references}
\clearpage
\label{appendices-start}
\input{sections/supplement-guide}
\input{sections/ai-disclosure}
\appendix
\renewcommand*{\theHsection}{appendix.\Alph{section}}
\renewcommand*{\theHsubsection}{\theHsection.\arabic{subsection}}
\input{sections/appendix-e-authentication/authentication-instance}
\input{sections/appendix-a-code/ltc-appendix}

\input{sections/appendix-a-code/completion-proofs}

\input{sections/appendix-b-recovery/recovery-supplement}
\input{sections/appendix-a-code/family-and-testing}

\input{sections/appendix-b-carrier/arithmetic-carrier}

\input{sections/appendix-a-code/client-addressing}

\input{sections/appendix-b-recovery/collective-das-supplement}

\end{document}

%% file: sections/01-introduction/introduction.tex
\section{Introduction}
Each block in a blockchain has a compact header containing a commitment to its much larger data payload.
Light clients download only the headers to avoid the bandwidth and storage cost of full blocks.
Yet a malicious publisher may release a header while withholding part or all of the underlying data. Light clients need to check data availability without downloading the full block.

Data availability sampling (DAS) addresses this problem by combining erasure coding with random sampling~\cite{FraudDAS}.
A suitable erasure code adds redundancy so that a correctly encoded block remains recoverable even when a constant fraction of its symbols is missing.
Making such a block unrecoverable requires withholding many encoded symbols, so sufficiently many independent random queries detect this withholding with high probability.
A publisher may also answer some clients and ignore others, and a single client's small sample need not suffice to reconstruct the full block.
In collective DAS, each client checks a small sample, and an extraction algorithm pools their observations to recover the message~\cite{Foundations,FRIDA}.
For a sufficiently large set of clients sampling independently, it should be unlikely that they all accept yet extraction fails.

Standard erasure decoding, however, requires the observed values to agree with a codeword.
A plain Merkle opening authenticates a symbol at its position in the committed word, but a malicious publisher may commit to a word outside the code, leaving recovery with errors as well as erasures.
Coded Merkle trees~\cite{CMT} and two-dimensional Reed--Solomon DAS~\cite{FraudDAS} address encoding validity through fraud proofs of incorrect encoding.
Erasure code commitments instead enforce \emph{code binding}, which requires verified openings to agree with a codeword~\cite[Sections~1.2 and~4.1]{Foundations}.

FRIDA~\cite{FRIDA} obtains code binding for Reed--Solomon codes using FRI~\cite{FRI}, without a trusted setup; both its commitment and its opening replies carry proof data.\footnote{DEEP-FRIDA~\cite{DEEPFRIDA} shortens the commitment using list decoding, and Mohnblatt and Wagner~\cite{FRIDABeyond} analyze batched FRI beyond unique decoding. In a complementary research direction, DAS with Repair~\cite{DASRepair} studies local repair of stored shares and their opening proofs.}
For $K$ information symbols and security parameter~$\lambda$, its analysis~\cite[Section~5, Table~1]{FRIDA} reports $O(\lambda^2\log^2 K)$ commitment bits and $O(\lambda\log^2 K)$ bits per draw on the publisher--client hop.
Its chosen field bounds the evaluation domain, so these estimates apply to the input sizes that field supports~\cite[Section~4.1]{FRIDA}; longer blocks require a larger field.

The DAS construction of ZODA~\cite[Appendix~F]{ZODA} uses opened rows and columns for both checking and decoding.
It authenticates those lines and individual entries with three Merkle commitments, and its checks use random projections over a field of size greater than $2^\lambda$.
For a square message array of $K$ information symbols, each opened row or column contains $\sqrt{K}$ symbols.

These constructions motivate short samples of symbols whose bit width is independent of the block length and~$\lambda$.

\emph{Can collective DAS be achieved with a constant number of symbols per sample over a fixed finite field, a commitment consisting of one Merkle root, and opening replies containing only the sampled symbols and Merkle authentication data?}

Rather than require code binding from the commitment or rely on fraud proofs, we let each client test its sample against local parity equations, as in locally testable codes (LTCs).
A passing local test does not certify that the committed word is a codeword.
We therefore need the same local equations to support recovery from partial observations that may contain errors as well as erasures.
We call this \emph{checked recovery}: it returns a codeword only when the numbers of unobserved coordinates and of disagreements with the observations lie within prescribed budgets.

We seek constant-query LTCs over any fixed finite field at any target rate below~$1$, with fast systematic encoding. We also require checked recovery at every fixed positive erasure fraction below half the guaranteed relative distance.
High rate keeps redundancy low, while fast systematic encoding supports efficient publication and lets the extractor read the message directly from a recovered codeword.
Earlier LTC constructions establish different parts of this picture.
Dinur et al.~\cite[Theorem~1.2 and Algorithm~4.6]{DELLM} construct binary left--right LTCs at any target rate below~$1$ and give a linear-time decoder correcting a sufficiently small fraction of errors in fully observed binary words.
Panteleev and Kalachev~\cite[Theorem~1]{PK} construct good LTCs (with positive constant rate and relative distance) over any finite field for target rates below~$1/2$.
First and Lazarovici~\cite[Theorems~1(ii) and~3(ii)]{FirstLazarovici} establish the existence of good $3$-query linear LTCs over fixed finite fields.
Viderman's binary construction~\cite[Corollary~3.8]{Viderman} of length $N$ gives linear-time encoding and $N^\epsilon$-query testing for any fixed $\epsilon>0$.

\subsection{Our Contributions}
Inspired by the left--right code geometry of~\cite{DELLM,PK}, we construct a family of LTCs over any fixed finite field that combines constant-query local testing, fast systematic encoding onto the entire code defined by the line checks, and checked recovery from partial observations.
Because encoding covers the whole code, every word returned by recovery is the encoding of the message read from its information coordinates.
The same local equations thus govern publication, testing, and extraction, yielding collective DAS with a single Merkle commitment and no trusted setup.

The coordinates of our LTC lie on intersecting lines of fixed length $D$.
Each client draws one line label uniformly, authenticates its symbols against the commitment, and checks the line's parity equations.
Opening a line reveals the symbols at its crossings with other lines, supplying observations even on unsampled lines.
From the merged authenticated observations, the extractor runs checked recovery using the same parity equations and line overlaps, and outputs the returned codeword's information symbols whenever recovery succeeds.
The commitment and random-oracle log at registration fix the reference word for soundness. If the word fails many line checks, independent samples rarely all pass. If few checks fail, checked recovery succeeds once the pooled observations meet the coverage requirements (Section~\ref{sec:overview}).

\medskip
\noindent\textbf{1. Collective DAS.}
The scheme sets the recovery budget for unobserved coordinates as a fraction~$\zeta$ of the block length.
For sufficiently large members, $O_{\rm fam,\zeta}(N/D+\lambda)$ clients suffice for collective DAS.\footnote{Subscripts on $O$ and $\Theta$ list the fixed parameters on which hidden constants may depend; $\mathrm{fam}$ denotes the fixed local and arithmetic parameters of the code family. The hidden constants are independent of $N$ and~$\lambda$.}
Each client samples and authenticates one $D$-symbol line under a shared Merkle root, giving the coverage scale $N/D$.
Under the conditions of Theorem~\ref{thm:main-das}, the completeness, soundness, and consistency errors are at most $2^{-\lambda}$ in the classical random-oracle model with adaptive publisher replies.

\medskip
\noindent\textbf{2. Encoding and recovery on one LTC family.}
Proposition~\ref{prop:family} establishes a uniform left--right family over any fixed finite field with any target rate below~$1$ and positive constant relative distance, with local parity-check matrices selected once in public preprocessing.
The same line equations support constant-query local testing on full words (Theorem~\ref{thm:testing}) and checked algebraic recovery from partial observations, which succeeds under the local-check and erasure conditions of Theorem~\ref{thm:recovery}.
After reusable public preprocessing, systematic encoding onto the full code takes $O_{\rm fam}(N\log N)$ operations in the code's finite field and on indices, and algebraic recovery takes $O_{\rm fam}(N)$ such operations on any merged partial word (Proposition~\ref{prop:recovery-cost}).

\medskip
\noindent\textbf{3. Communication.}
For a family member with $K$ information symbols, constant rate gives $N=\Theta_{\rm fam}(K)$.
Our commitment digest uses $O(\lambda)$ bits when the total random-oracle query budget across a security experiment is bounded by $2^{O(\lambda)}$ and the digest width is set at the threshold prescribed in Theorem~\ref{thm:main-das}.
Under these conditions, a line request together with its Merkle-authenticated reply carrying $D$ field symbols uses $O_{\rm fam}(\lambda\log K)$ bits on the publisher--client hop.
For comparison, FRIDA also has an $O(K+\lambda)$ sufficient-draw bound; Table~\ref{tab:sampling-costs} (Section~\ref{sec:family-resources}) compares the commitment and per-draw publisher--client costs.
Appendix~\ref{app:full-resource-ledger} gives our full resource bounds.

\subsection{Main Theorem}
Our experiments (Section~\ref{sec:model}) adapt the completeness, soundness, and consistency notions of~\cite[Definition~1]{Foundations} to message extraction.
The publisher registers a pair consisting of a commitment and a public context identifying the code and protocol parameters, and may adapt its replies to the clients' samples.

The experiments impose fixed public resource bounds, including limits on client sets and transcript lists and a cap $\Qhash\ge1$ on all random-oracle queries. A client set or transcript list is \emph{supported} if it meets the corresponding bounds.

\begingroup
\displaywidowpenalty=\widowpenalty
\begin{theorem}\label{thm:main-das}
For every finite field $\F$ of characteristic $p$ and every target rate $\rho\in(0,1)$, there exist a fixed line length $D$, a constant $\Delta>0$, and a family of codes over $\F$ indexed by the nonnegative integers $m$. Member $m$ has length $N=D^2p^m$, rate at least $\rho$, and relative distance at least $\Delta$. All members use the same local parity-check matrices.

For every $0<\zeta<\Delta/2$, each sufficiently large member gives a collective DAS scheme with a cap $u$ on unobserved coordinates and a disagreement budget $t$ satisfying
\[
 u=\lfloor\zeta N\rfloor,\qquad t=\Theta_{\rm fam,\zeta}(N),\qquad 2(t+u)<\lceil\Delta N\rceil.
\]
The recovery budgets and the threshold on code length are independent of the security parameter. For each such member and integer $\lambda\ge1$, there is a sufficient client threshold $n_{\rm req}=O_{\rm fam,\zeta}(N/D+\lambda)$.

Fix the public resource bounds and the Merkle digest width $\kappa$ before choosing a message or running the adversary. The bounds must permit a complete honest run with at least $n_{\rm req}$ clients. Choose $\kappa$ at or above the threshold in~\eqref{eq:fixed-das-digest-width} of Section~\ref{sec:family-guarantee}, which is $\lambda+2\log_2\Qhash+O(1)$ bits.

The following guarantees hold in the classical random-oracle model for experiments within these public bounds. For completeness and soundness, any rule may select a supported set of $\ell\ge n_{\rm req}$ clients after registration, possibly using the registration history. Each selected client then draws one uniform line label using fresh, independent randomness.
\begin{enumerate}
\item \emph{Completeness.} For every message, honest publication and complete honest replies make every selected client accept. Over the clients' random draws, extraction from their transcript list returns the published message with probability at least $1-2^{-\lambda}$.
\item \emph{Soundness.} For every adversary that registers the commitment and context and controls the replies, the joint probability that all selected clients accept and extraction from their list returns $\bot$ is at most $2^{-\lambda}$.
\item \emph{Consistency.} For every adversary that supplies one registered pair and two supported transcript lists, the probability that both extractions succeed and return different messages is at most $2^{-\lambda}$.
\end{enumerate}
\end{theorem}
\endgroup

The \hyperref[proof:main-das-composition]{proof in Section~\ref*{sec:family-guarantee}} applies Theorem~\ref{thm:das} to the fixed-field family.
The consistency proof uses Merkle position binding to obtain agreement on shared observations under the same registered pair.
Combined with the verified erasure and disagreement bounds ($u$ and $t$), the distance margin then forces any two successfully recovered codewords to coincide (Section~\ref{sec:instance-das}).

%% file: sections/02-overview/overview.tex
\section{Technical Overview}\label{sec:overview}
The code $\C$ is defined by parity equations on intersecting $D$-symbol lines. Opening one line supplies observations on other lines, and the same equations guide recovery from those observations. Since the systematic encoder maps messages bijectively onto all of $\C$, the extractor can read the message from the information coordinates of any recovered codeword.
The \hyperref[app:common-notation]{supplementary notation table} collects the symbols used below.

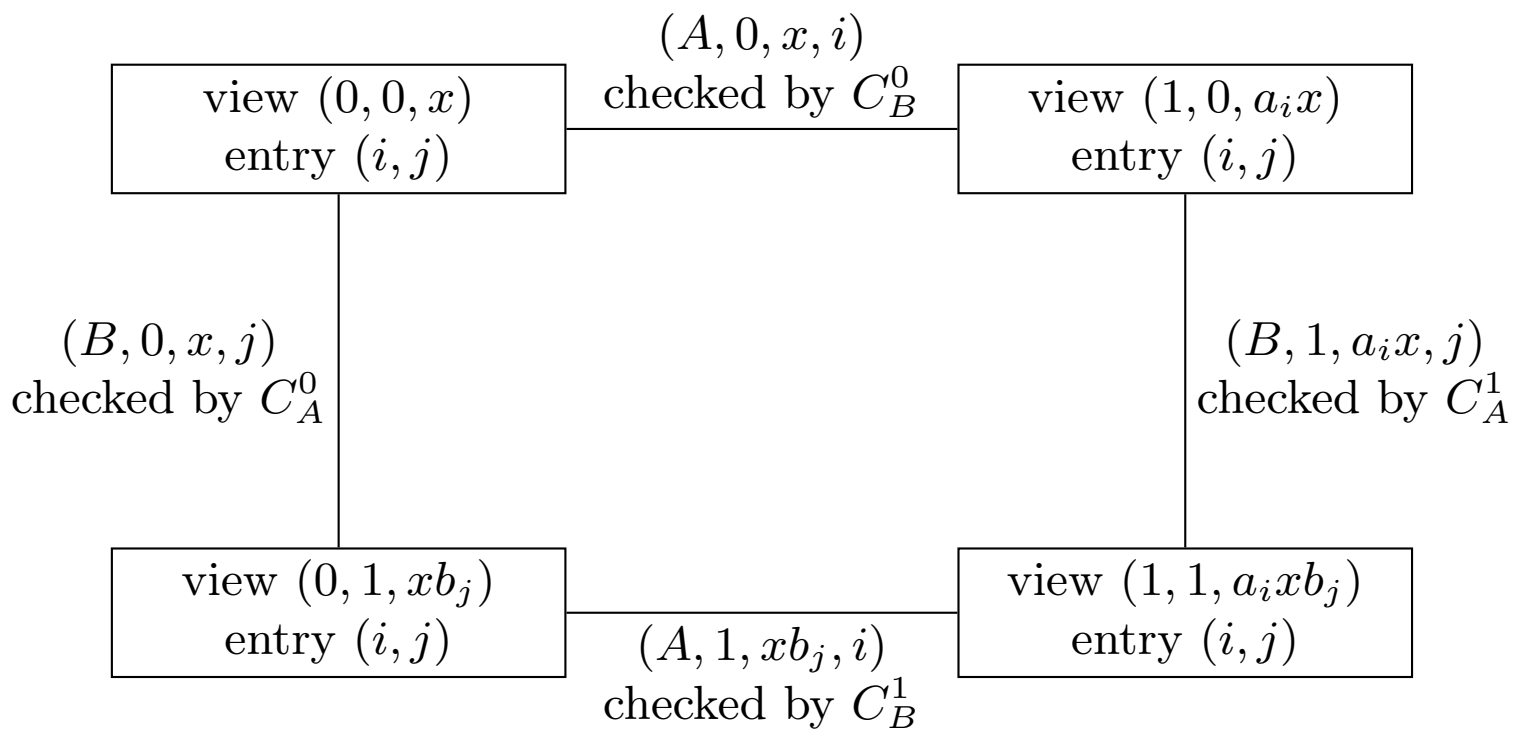
\begin{figure}[htb]
\centering
\begin{tikzpicture}[x=1cm,y=1cm,font=\small,every node/.style={inner sep=2pt}]
\tikzset{view/.style={draw,minimum width=3.0cm,minimum height=.85cm,align=center}}
\node[view] (tl) at (0,3.2) {view $(0,0,x)$\\ entry $(i,j)$};
\node[view] (tr) at (5.6,3.2) {view $(1,0,a_ix)$\\ entry $(i,j)$};
\node[view] (bl) at (0,0) {view $(0,1,xb_j)$\\ entry $(i,j)$};
\node[view] (br) at (5.6,0) {view $(1,1,a_ixb_j)$\\ entry $(i,j)$};
\draw (tl) -- node[above,align=center]{$(A,0,x,i)$\\ checked by $C_B^0$} (tr);
\draw (bl) -- node[below,align=center]{$(A,1,xb_j,i)$\\ checked by $C_B^1$} (br);
\draw (tl) -- node[left,align=center]{$(B,0,x,j)$\\ checked by $C_A^0$} (bl);
\draw (tr) -- node[right,align=center]{$(B,1,a_ix,j)$\\ checked by $C_A^1$} (br);
\end{tikzpicture}
\caption{Four views of one coordinate $(x,i,j)\in G\times[D]^2$, where $G$ is the indexing group. A view label gives two type bits $s_A,s_B$ and a group element. An $A$-edge translates by $a_i$ on the left and fixes row $i$, so its entries vary over $j$ and are checked by $C_B^{s_B}$. A $B$-edge translates by $b_j$ on the right and fixes column $j$, so its entries vary over $i$ and are checked by $C_A^{s_A}$ (Section~\ref{sec:source-hypotheses}). Opening any of these four lines reveals the symbol at $(x,i,j)$.}
\label{fig:local-square}
\end{figure}

\paragraph{\textup{\textbf{Shared views and lines.}}}
A \emph{view} is a $D\times D$ matrix. Each row or column is a line shared by two views. A view is legal if its rows and columns pass their respective checks, and a word belongs to $\C$ if all its views are legal.
Figure~\ref{fig:local-square} shows the four views and four lines that share a coordinate.

\paragraph{\textup{\textbf{From samples to soundness.}}}
The commitment and oracle log at registration fix a reference word $W\in\F^N$, which may violate the line checks. Outside the fixation error event, verified openings agree with $W$ at their coordinates (Section~\ref{sec:auth-interface}). The extractor retains symbols from complete replies that pass transcript validation, even if a sampled line fails its parity check (Section~\ref{sec:protocol}).

The registration history determines whether $W$ exceeds the public budget $\ill_{\max}$ of failed line checks before the clients sample. If it does, unanimous acceptance outside the fixation error event requires every draw to miss the failed lines. Otherwise, unanimous acceptance outside that event gives the extractor $W|_{S^*}$ on the requested coordinates $S^*$. Few failed checks limit the views initially left unassigned because of incompatible observations. Sufficient coverage bounds the total erasure count and the number of \emph{heavy} lines, those with too many erasures. With admissible parameters, these two bounds suffice for recovery in this case.
The joint probability of unanimous acceptance and extraction failure is thus at most the fixation error plus the larger of the two sampling error bounds, one for missing the failed lines and one for insufficient coverage (Theorem~\ref{thm:das}).

\paragraph{\textup{\textbf{From observations to a message.}}}
Given observations $y\in\F^S$ on the observed set $S$, recovery constructs and checks a candidate codeword through initialization, completion, and verification (Figure~\ref{fig:recovery-algorithm} gives the full algorithm). In the recovery case above, $S=S^*$.

\emph{Initialization.}
Recovery completes partially observed lines where possible and uses those completions to assign legal views that agree along shared lines. An unassigned view is a \emph{hole}. Views touching heavy lines become holes. Every other initial hole is incident to several lines that fail their checks in each full word extending $y$. The budgets on heavy and failed lines therefore control the initial hole set (Lemma~\ref{lem:recovery-initialization} and Section~\ref{sec:recovery-parameters}).

\emph{Completion.}
Closure enlarges the hole set until each assigned view shares only a few labelled lines with holes. For a small initial hole set, expansion bounds this enlargement and ensures that, whenever holes remain, some hole has enough lines supplied by assigned neighbours. Two supplied lines can disagree only at crossings where the view opposite the hole is also a hole (Figure~\ref{fig:local-square}). That opposite view neighbours both suppliers, so the bound on each supplier's edges to holes limits these crossings on each supplied line. The local codes' sparse-agreement property (Definition~\ref{ltc:je}) then forces agreement. This lets recovery fill the hole while preserving its supplied lines (Lemma~\ref{lem:hole-completion} and Section~\ref{sec:shared-completion}).

\emph{Verification.}
Recovery merges the completed views into a word $c$ and checks $c\in\C$, $N-|S|\le u$, and $\wt(c|_S-y)\le t$, where $u$ bounds erasures and $t$ bounds disagreements on observed coordinates. It returns $c$ only after these checks pass, and the extractor reads its information symbols. Theorem~\ref{thm:recovery} gives sufficient conditions for recovery to succeed with the admissible budgets of Section~\ref{sec:recovery-parameters}.

\paragraph{\textup{\textbf{Completeness and consistency.}}}
An honest word passes every line check, and with complete replies and sufficient coverage recovery succeeds. Its output differs from the published word in at most $t+u$ coordinates, so the margin $2(t+u)<d_C$, with $d_C$ a lower bound on the minimum distance of $\C$, ensures that the output is the published codeword.
For consistency, position binding gives agreement on shared coordinates outside the binding error event. The same distance margin forces two successful outputs under one registration to coincide (Section~\ref{sec:instance-das}).

\paragraph{\textup{\textbf{Systematic encoding and the fixed-field family.}}}
The encoder solves the left and right line equations successively, keeping the information coordinates fixed (Section~\ref{sec:systematic-encoding}). An arithmetic tower lets the block length grow while the code field and line length stay fixed. The tower supplies both the expansion used for distance and hole closure and the central quotients used for fast encoding (Section~\ref{sec:groups-family}).

%% file: sections/03-model/model.tex
\section{Collective DAS: Model, Interfaces, and Protocol}\label{sec:model}
We specify collective DAS using a general line-recovery code and authentication layer, allowing one or more line draws per client.
We fix the code, message space, recovery budgets, minimum client count $\ell_0$, sampling rule, resource bounds, and authentication parameters before message choice, and complete public preparation before registration.
The publisher registers a pair $(\mathsf{ctx},\mathsf{com})$ consisting of a public context and a commitment.
Clients then sample and decide locally whether to accept, while the publisher may adapt, withhold, or interleave replies.

A client transcript records the drawn line labels and the reply, if any.
The deterministic extractor $\Ext$ takes a registered pair and a list of transcripts and returns a message or $\bot$ (Section~\ref{sec:protocol}).
Following~\cite[Definition~1]{Foundations} and~\cite[Appendix~A, Definition~14]{FRIDA}, the definition below states the standard completeness, soundness, and consistency events with concrete error bounds.
Section~\ref{sec:generic-security} proves these guarantees for the protocol.

\begin{definition}[Collective DAS security]\label{def:das}
The error bounds apply to the specified adversary class in complete experiments within the public resource bounds.
A client population or transcript list is \emph{supported} if it meets the corresponding public resource bounds, including its allowed size. A supported list may contain entries that fail validation. The public resource bounds account for setup, preparation, execution including failed or repeated attempts, and verification by clients and the recovery party. The total oracle-query budget counts queries by the adversary and by all algorithms in the experiment. Probabilities are over the randomness of setup, the oracle when used, the adversary, and any client selection and draws.

For completeness and soundness, let a selection rule select any supported set $\mathcal I$ of $\ell\ge\ell_0$ clients after registration.
The rule may depend on the registration history specified in Section~\ref{sec:auth-interface}.
The clients in $\mathcal I$ then draw fresh, independent sampling coins, and extraction uses their transcript list.
\begin{enumerate}
\item \emph{Completeness.} For every message, honest publication and complete honest replies make every client in $\mathcal I$ accept. Extraction returns the published message with probability at least $1-\eps_{\rm comp}(\ell)$.
\item \emph{Soundness.} The adversary registers the pair and controls the replies. The joint probability that every client in $\mathcal I$ accepts and $\Ext$ returns $\bot$ is at most $\eps_{\rm snd}(\ell)$.
\item \emph{Consistency.} The adversary supplies one registered pair and two supported transcript lists. The client minimum $\ell_0$ applies only to completeness and soundness. The probability that the two invocations of $\Ext$ both return messages and those messages differ is at most $\eps_{\rm cons}$.
\end{enumerate}
For each fixed $\ell\ge\ell_0$, the completeness and soundness bounds must hold uniformly over supported selection rules of size $\ell$.
For supported selection rules with variable $\ell\ge\ell_0$, the error bounds are $\eps_{\rm comp}(\ell_0)$ and $\eps_{\rm snd}(\ell_0)$.
\end{definition}

\input{sections/03-model/notation}

\subsection{DAS protocol}
\label{sec:protocol}

The \emph{recovery party} runs the deterministic extractor $\Ext$ on the client transcripts to recover the message or return $\bot$.

The authentication layer provides commitment and opening algorithms, a canonical reply format, and a deterministic verifier.
Given a transcript containing draw labels and a purported reply, the verifier checks the registered pair $(\mathsf{ctx},\mathsf{com})$, the requested coordinates, the values, and the authentication data.
A complete canonical reply that passes these checks is \emph{authenticated}, and honest replies verify.

The public parameters specify the line-recovery code, information order, field representation, serialization, coordinate numbering, and per-client draw count $a_{\rm cl}$. Clients and the extractor check that the registered context $\mathsf{ctx}$ matches the context determined by these public parameters.

\paragraph{Publication and queries.}
For a message $\mathbf m\in\F^K$, an honest publisher computes $\mathsf{Enc}_{\C}(\mathbf m)$, commits to the resulting codeword, and registers $(\mathsf{ctx},\mathsf{com})$.
After registration, a selection rule selects a supported set $\mathcal I$ of $\ell$ clients indexed by $i\in[\ell]$.
Each client then independently draws $a_{\rm cl}\ge1$ uniformly random line labels with replacement.

The requested coordinate set for client $i$ is
\[
 S_i^*=\bigcup_{\mathcal L\text{ drawn by client }i}\mathcal L.
\]
Write $S^*=\bigcup_i S_i^*$ for the \emph{requested coordinate set} determined by the selected clients' draws.

An honest publisher returns every requested value with its authentication opening.
Client $i$ checks the registered context and verifies that the reply is complete, canonical, and authenticated for the requested set $S_i^*$.
It then checks every parity equation on its drawn lines. The client accepts if and only if the context, reply-verification, and parity checks all pass.
The client forwards its draw labels to the recovery party in a transcript, together with the unchanged purported reply or a record that no reply was received.

\paragraph{Extraction.}
Given $(\mathsf{ctx},\mathsf{com})$ and a finite transcript list $\mathcal T$, $\Ext$ first checks the context, returning $\bot$ if it is unexpected.
For each transcript, it checks the draw labels and requested coordinates and verifies that the reply is present, complete, canonical, and authenticated.
It discards entries that fail any of these checks and retains the authenticated values from every entry that passes, even when a sampled line fails its parity check.
It merges the retained values into a partial word $y\in\F^S$ on the observed set $S$, returning $\bot$ if two retained values conflict at a shared coordinate.
It then invokes $\mathsf{Rec}$ on $y$.
The recovery algorithm checks the output conditions of Definition~\ref{def:line-recovery} before returning a codeword (Section~\ref{sec:recovery-algorithm}).
The extractor returns the codeword's information symbols in their prescribed order, or $\bot$ if recovery fails.

%% file: sections/03-model/notation.tex
\subsection{Code and recovery interface}
\label{sec:code-interface}

We write $[n]=\{0,\ldots,n-1\}$, $\F^S$ for words on an index set $S$ (with $\F^n=\F^{[n]}$), $W|_S$ for restriction, $C|_S$ for puncturing, $\wt$ for Hamming weight, and $\dist$ for relative Hamming distance. The code alphabet is a finite field $\F$ of size $q$. Unless stated otherwise, logarithms in probability estimates are natural.

The interface specifies a systematic linear code $\C\subseteq\F^N$ of dimension $K$, a public encoder $\mathsf{Enc}_{\C}$, a fixed ordered information set, and $M$ labelled parity-check lines.
Restriction to the information set inverts the encoder.
Each label specifies $D$ distinct coordinates in a public order and the parity equations to check there.
For $W\in\F^N$, write $\ill(W)$ for the number of illegal lines in $W$, namely labels on which a check fails.
We require the line checks to define $\C$, so $W\in\C$ exactly when $\ill(W)=0$.
Write $d_C$ for a positive integer lower bound on the minimum distance.
With a relative distance bound $\Delta$, take $d_C=\lceil N\Delta\rceil$.

In set operations, a line label $\mathcal L$ denotes its set of $D$ physical coordinates.
For a public integer threshold $h\ge 0$ and $S\subseteq[N]$, a line is erasure-heavy if $|\mathcal L\setminus S|>h$.
Write $r_{\rm H}(S)$ for the number of erasure-heavy line labels, counting distinct labels separately even if they have the same coordinate set.

The recovery guarantee uses budgets for illegal lines, heavy lines, and missing coordinates. Section~\ref{sec:recovery} constructs an algorithm meeting this interface.

\begin{definition}[Line-recovery code]\label{def:line-recovery}
A \emph{line-recovery code} consists of a code as specified above, public integers $h,\ill_{\max},R,u,t\ge0$ with $\ill_{\max}<M$, $u<N$, and $2(t+u)<d_C$, and a public deterministic recovery algorithm $\mathsf{Rec}$ that, on every $S\subseteq[N]$ and $y\in\F^S$, returns a word of $\F^N$ or $\bot$.
With $r_{\rm H}$ taken at threshold $h$, $\mathsf{Rec}$ has the following two properties.
\begin{enumerate}
\item \emph{Recovery under the budgets.}
For every $W\in\F^N$ and $S\subseteq[N]$ with $\ill(W)\le\ill_{\max}$, $r_{\rm H}(S)\le R$, and $N-|S|\le u$, $\mathsf{Rec}$ returns a codeword when invoked on $W|_S$.
\item \emph{Checked outputs.}
If $\mathsf{Rec}$ returns $c$ on $y\in\F^S$, then $c\in\C$, $N-|S|\le u$, and $\wt(c|_S-y)\le t$.
\end{enumerate}
\end{definition}

\begin{lemma}\label{lem:interface-uniqueness}
Let $\mathsf{Rec}$ be any partial-word algorithm satisfying the checked-output property above for a code of minimum distance at least $d_C$, with budgets $t,u$ satisfying $2(t+u)<d_C$.
If $\mathsf{Rec}$ returns $c_1$ on $y_1\in\F^{S_1}$ and $c_2$ on $y_2\in\F^{S_2}$, where $y_1$ and $y_2$ agree on $S_1\cap S_2$, then $c_1=c_2$.
In particular, if $\mathsf{Rec}$ returns $c$ on $W|_S$ for some $W\in\C$, then $c=W$.
\end{lemma}

\begin{proof}
At each position in $S_1\cap S_2$ where $c_1$ and $c_2$ differ, at least one of them disagrees with its corresponding observation.
Outside $S_1\cap S_2$, at most $N-|S_1\cap S_2|\le2u$ disagreements are possible.
The checked-output bounds therefore give
\[
 \wt(c_1-c_2)\le\sum_{i=1}^2\wt(c_i|_{S_i}-y_i)+N-|S_1\cap S_2|\le2(t+u)<d_C.
\]
Since $c_1,c_2\in\C$, the distance bound gives $c_1=c_2$. For $W\in\C$, checked outputs give $\wt(c-W)\le\wt(c|_S-W|_S)+(N-|S|)\le t+u<d_C$, forcing $c=W$.
\end{proof}

%% file: sections/03-model/generic-security.tex
\section{Generic Security of Collective DAS}\label{sec:generic-security}
We combine authentication with the recovery interface to prove the protocol's completeness, soundness, and consistency.
Fixation will connect the observed values to a word fixed before sampling in the soundness proof, while position binding will relate the observations used by two extractions in the consistency proof.

\subsection{Authentication guarantees}
\label{sec:auth-interface}

\paragraph{Fixation.}
Let $\mathsf{hist}$ denote the complete history at registration of $(\mathsf{ctx},\mathsf{com})$, including all oracle queries and answers up to that point, before the clients draw their sampling coins.
The authentication layer specifies a function mapping $\mathsf{hist}$ to a reference word $W(\mathsf{hist})\in\F^N$, so the lines on which $W$ violates the local checks are fixed before sampling.
Consider the continuation with a supported set of $\ell\ge\ell_0$ clients selected as in Definition~\ref{def:das}.
Let $\Badval$ be the event that a client or the recovery party authenticates, under the registered pair, a reply that assigns some coordinate a value different from $W$.
The authentication layer guarantees $\Pr[\Badval]\le\eps_{\rm fix}$ over the complete experiment of Definition~\ref{def:das}, including adaptive publisher replies.

\paragraph{Position binding.}
In a separate experiment, the adversary supplies one registered pair and two arbitrary finite lists of coordinate-labelled replies.
In particular, a transcript from Section~\ref{sec:protocol} carrying a reply gives such a coordinate-labelled reply, with its draw labels determining the requested coordinates.
The experiment is subject to the complete-experiment resource bounds supported by the authentication layer, which account for verification by the recovery party as in Definition~\ref{def:das}.
The event $\mathsf{Bad}_{\rm bind}$ occurs when two of the supplied replies pass authentication verification under this pair but assign different values to a shared coordinate.
The authentication layer guarantees $\Pr[\mathsf{Bad}_{\rm bind}]\le\eps_{\rm bind}$.

\paragraph{Merkle instance.}
A single Merkle tree with one leaf per physical coordinate supplies both guarantees.
Appendix~\ref{app:authentication-instance} specifies the tree and context binding. Lemma~\ref{lem:auth-instance} gives the fixation and position-binding bounds for the classical random-oracle model under the complete-experiment query cap $\Qhash$, including for computationally unbounded adversaries.

\input{sections/03-model/das-composition}

%% file: sections/03-model/das-composition.tex
\subsection{Security composition}\label{sec:instance-das}

The soundness proof uses two sampling bounds: one for insufficient coverage, and one for avoiding all illegal lines.

After registration, let a selection rule select any supported client set $\mathcal I$ with $\ell=|\mathcal I|\ge\ell_0$.
As in Section~\ref{sec:protocol}, the clients in $\mathcal I$ then make $n_{\rm dr}=a_{\rm cl}\ell$ independent uniform line draws, determining the requested coordinate set $S^*=\bigcup_i S_i^*$.
Put $V=N-|S^*|$ and use the heavy-line count $r_{\rm H}(S^*)$ of Section~\ref{sec:code-interface}.
Given a registration history $\mathsf{hist}$ and the selected set $\mathcal I$, the observation-failure condition over the clients' fresh sampling coins is
\begin{equation}\label{eq:observation-failure}
 \Pr[V>u\ \text{or}\ r_{\rm H}(S^*)>R\mid\mathsf{hist},\mathcal I]\le\eps_{\rm obs}.
\end{equation}
With the public parameters fixed, the sampling distribution depends on $\mathcal I$ only through $\ell$.
Couple the draws at size $\ell_0$ with the first $a_{\rm cl}\ell_0$ draws at any size $\ell\ge\ell_0$.
The remaining draws can only enlarge $S^*$, so both $V$ and $r_{\rm H}(S^*)$ are nonincreasing. Thus the same bound $\eps_{\rm obs}$ verified at $\ell_0$ holds for every $\ell\ge\ell_0$.
The fixed-family instantiation in Section~\ref{sec:family-guarantee} gives a draw count satisfying~\eqref{eq:observation-failure}.

For any fixed word $W$ with more than $\ill_{\max}$ illegal labelled lines, the probability that all $n_{\rm dr}$ independent draws avoid those lines is at most
\begin{equation}\label{eq:lt-prob}
 \eps_{\rm avoid}(n_{\rm dr})=(1-(\ill_{\max}+1)/M)^{n_{\rm dr}}.
\end{equation}
The bound is nonincreasing in $n_{\rm dr}$.

\begin{theorem}
\label{thm:das}
Fix a line-recovery code (Definition~\ref{def:line-recovery}) and an authentication layer as in Section~\ref{sec:protocol} satisfying the security guarantees of Section~\ref{sec:auth-interface}. Clients request exactly their drawn lines and forward their transcripts as in Section~\ref{sec:protocol}.

Suppose~\eqref{eq:observation-failure} holds for every registration history and every supported set $\mathcal I$ of $\ell\ge\ell_0$ clients selected as in Definition~\ref{def:das}.
The adversary class and complete-experiment resource bounds are those supported by the authentication layer.

Within these resource bounds, the protocol satisfies Definition~\ref{def:das} with
\[
 \eps_{\rm comp}(\ell)\le\eps_{\rm obs},\qquad
 \eps_{\rm snd}(\ell)
   \le\max\{\eps_{\rm avoid}(a_{\rm cl}\ell),\eps_{\rm obs}\}+\eps_{\rm fix},
 \qquad
 \eps_{\rm cons}\le\eps_{\rm bind}.
\]
For a variable-size rule, use the bounds at $\ell_0$.
\end{theorem}

\begin{proof}
\textbf{Completeness.}
For an honestly published message $\mathbf m$, the systematic encoder produces $W_{\rm hon}=\mathsf{Enc}_{\C}(\mathbf m)\in\C$, so $\ill(W_{\rm hon})=0\le \ill_{\max}$.
With complete honest replies, all authentication and line checks pass, every selected client accepts and forwards its transcript, and $\Ext$ retains every selected client's transcript.
The extractor therefore obtains $y=W_{\rm hon}|_{S^*}$ with $S=S^*$.
Whenever $V\le u$ and $r_{\rm H}(S^*)\le R$, the recovery guarantee makes $\mathsf{Rec}$ return a word $c$, and Lemma~\ref{lem:interface-uniqueness} gives $c=W_{\rm hon}$.
Extracting its systematic coordinates returns $\mathbf m$.
Equation~\eqref{eq:observation-failure} thus gives $\eps_{\rm comp}(\ell)\le\eps_{\rm obs}$.

\textbf{Soundness.}
Let $\mathsf{Fail}$ be the joint event that every selected client accepts and $\Ext$ returns $\bot$ on their transcript list.
Condition on the registration history $\mathsf{hist}$ and the selected set $\mathcal I$.
Under this conditioning, $W=W(\mathsf{hist})$ and its illegal lines are fixed, and the clients' draws are independent and uniform.
On unanimous acceptance, every selected client forwards its complete authenticated transcript unchanged, so it passes $\Ext$'s deterministic checks under the same registered pair.
Outside $\Badval$, these values agree with $W$, so $\Ext$ retains all their transcripts and merges $y=W|_{S^*}$, with $S=S^*$.

If $\ill(W)>\ill_{\max}$, outside $\Badval$ a complete reply for any drawn illegal line fails the parity check.
Thus $\mathsf{Fail}\cap\Badval^{\,c}$ requires all $a_{\rm cl}|\mathcal I|$ draws to avoid illegal lines, an event of conditional probability at most $\eps_{\rm avoid}(a_{\rm cl}|\mathcal I|)$ by~\eqref{eq:lt-prob}.

If $\ill(W)\le\ill_{\max}$, recovery returns a codeword from $W|_{S^*}$ when $V\le u$ and $r_{\rm H}(S^*)\le R$ (Definition~\ref{def:line-recovery}).
The extractor then reads its systematic message, so $\mathsf{Fail}\cap\Badval^{\,c}$ requires the event in~\eqref{eq:observation-failure}, of conditional probability at most $\eps_{\rm obs}$.

The history fixes which case applies, giving
\[
 \Pr[\mathsf{Fail}\cap\Badval^{\,c}\mid\mathsf{hist},\mathcal I]
 \le\max\{\eps_{\rm avoid}(a_{\rm cl}|\mathcal I|),\eps_{\rm obs}\}.
\]
For a variable-size rule, monotonicity bounds this by its value at $|\mathcal I|=\ell_0$.
Averaging and adding $\Pr[\Badval]\le\eps_{\rm fix}$ gives the joint soundness bound.

\textbf{Consistency.}
Consider any two supported transcript lists under the same registered pair $(\mathsf{ctx},\mathsf{com})$.
If both extractions succeed and return different messages, the two recovered codewords differ. Lemma~\ref{lem:interface-uniqueness} then implies that the two merged partial words disagree at a shared coordinate.
Those two values come from replies that passed the extractor's authentication checks under the registered pair, so the disagreement witnesses $\mathsf{Bad}_{\rm bind}$. The extractor's verification is part of the same complete experiment. The consistency error is therefore at most $\eps_{\rm bind}$.
\end{proof}

Extraction can also succeed when some clients reject. Outside $\Badval$, complete authenticated replies under the expected context agree with $W$, even on lines that fail their checks. If these replies cover $S^*$ and $W$ and $S^*$ meet the budgets of Definition~\ref{def:line-recovery}, recovery succeeds from $y=W|_{S^*}$.

Sections~\ref{sec:ltc}--\ref{sec:recovery} construct the required code and recovery algorithm, and Section~\ref{sec:security} instantiates them over a fixed finite field to prove Theorem~\ref{thm:main-das}.

%% file: sections/04-code/ltc.tex
\section{Line-Defined Codes and Systematic Encoding}\label{ltc:sec}\label{sec:ltc}

The left--right tensor-line geometry of~\cite{DELLM,PK} supplies the labelled lines for client queries.
The code is the common kernel of the line checks, and its systematic encoder is a bijection onto this kernel.
We first give its construction, systematic encoder, and distance bounds. Section~\ref{sec:shared-completion} develops the completion tool used for testing and recovery.

\subsection{Code construction and the encoding theorem}\label{sec:source-hypotheses}

Let $\F$ be a finite field of characteristic $p$, and let $G$ be a finite $p$-group of order $g=p^m$.
Fix two lists of $D$ labelled group elements (ports), $(a_i)_{i\in[D]}$ and $(b_j)_{j\in[D]}$, for left and right translation.
The lists may include the identity, and repeated group elements are unavoidable when $g<D$.
Port labels distinguish lines and view-graph edges, and all degree and edge counts include this multiplicity.

For $X\in\{A,B\}$ and $s\in\{0,1\}$, fix parity-check matrices $H_X^s\in\F^{r_X\times D}$, where $0<2r_X<D$.
Require the stacked matrix $[H_X^0;H_X^1]\in\F^{2r_X\times D}$, with the rows of $H_X^0$ above those of $H_X^1$, to have full row rank, and put
\[
 C_X^s=\ker H_X^s,\quad C_X^{\cap}=\ker[H_X^0;H_X^1],\quad k_X=\dim C_X^{\cap}=D-2r_X.
\]
The physical coordinates are $(x,i,j)\in G\times[D]^2$, giving block length $N=D^2g$.
A local view is indexed by $z=(s_A,s_B,x)$, where $s_A,s_B\in\{0,1\}$ are type bits and $x\in G$.
We use inverse port shifts so that adjacent views share a row or column.
For $W\in\F^{G\times[D]^2}$, the $D\times D$ matrix $W_z$ has entry $W(a_i^{-s_A}xb_j^{-s_B},i,j)$ at position $(i,j)$.
For every $j$, views $(0,s_B,x)$ and $(1,s_B,a_ix)$ read the same coordinate $(xb_j^{-s_B},i,j)$ in row $i$ because $a_i^{-1}(a_ix)=x$.
Right translation by $b_j$ similarly pairs column $j$ of $(s_A,0,x)$ and $(s_A,1,xb_j)$.

A view of type $(s_A,s_B)$ is legal if it lies in $C_A^{s_A}\otimes C_B^{s_B}$: its columns lie in $C_A^{s_A}$ and its rows in $C_B^{s_B}$. The code $\C$ consists of words $W\in\F^{G\times[D]^2}$ with every view $W_z$ legal.
For $G=\{1\}$, all four view matrices coincide on one grid, giving $\C=C_A^{\cap}\otimes C_B^{\cap}$.

The view graph has $4g$ vertices and $M=4Dg$ labelled edges, with one edge for each line shared by two views. Views with even $s_A+s_B$ form side $0$, and those with odd $s_A+s_B$ form side $1$. Each edge flips one type bit, and its label identifies the translation direction.
A \emph{row line} has label $(A,s_B,x,i)$. Left translation by $a_i$ connects $(0,s_B,x)$ to $(1,s_B,a_ix)$ along row $i$, indexed by the $B$-ports and checked by $C_B^{s_B}$.
A \emph{column line} has label $(B,s_A,x,j)$. Right translation by $b_j$ connects $(s_A,0,x)$ to $(s_A,1,xb_j)$ along column $j$, indexed by the $A$-ports and checked by $C_A^{s_A}$.

The ordered physical-coordinate lists are
\[
 (A,s_B,x,i):\ ((xb_j^{-s_B},i,j))_{j\in[D]},\qquad
 (B,s_A,x,j):\ ((a_i^{-s_A}x,i,j))_{i\in[D]}.
\]
Every line has $D$ distinct coordinates. For each fixed $s_B$, row lines partition the coordinates, as do column lines for each fixed $s_A$. Thus each coordinate belongs to exactly four labelled lines, two of each kind (Figure~\ref{fig:local-square}). A word lies in $\C$ exactly when all these line checks hold. DAS clients sample these same lines.

Each physical coordinate $(x,i,j)$ has four occurrences, its \emph{corners}, at position $(i,j)$ of the views $(0,0,x)$, $(1,0,a_ix)$, $(0,1,xb_j)$, and $(1,1,a_ixb_j)$ (Figure~\ref{fig:local-square}). Relative to the view $(0,0,x)$, the last corner is its opposite view, and this relation extends symmetrically from each corner of the square. The four types are distinct, and their group coordinates may coincide.

The ordered coordinate lists specify the requests. Appendix~\ref{app:public-orders} gives the public numbering for serialization and Merkle leaves.
For encoding, assume a prepared central order-$p$ quotient chain with the data specified in Appendix~\ref{app:central-quotient-solve}.

\begin{theorem}\label{thm:encoding}
Assume the hypotheses of the code construction.
Then $\C$ is an $\F$-linear systematic code of length $N=D^2g$ and dimension $K=k_Ak_Bg$.
There is an explicit set of $K$ information coordinates, defined in Section~\ref{sec:systematic-encoding}, such that every assignment to them extends uniquely to a codeword of $\C$.
Given the group data, encoding uses $O_{D,p}(N(1+\log g))$ code-field, array, and $O(\log N)$-bit index operations.
\end{theorem}

\subsection{Systematic encoding by central quotients}\label{sec:systematic-encoding}
We express the code in the group algebra $\mathcal R=\F[G]$~\cite{Berman,Assmus}, defining $\mathbf X\in\mathcal R^{D\times D}$ by $\mathbf X_{ij}=\sum_{x\in G}W(x,i,j)x$. Left and right translations commute. The defining checks are precisely $\mathbf A\mathbf X=0$ and $\mathbf X\mathbf B=0$, where
\[
\begin{aligned}
\mathbf A[(s_A,\nu),i]&=H_A^{s_A}[\nu,i]a_i^{s_A} &&(\nu\in[r_A]),\\
\mathbf B[j,(s_B,\nu)]&=H_B^{s_B}[\nu,j]b_j^{s_B} &&(\nu\in[r_B]).
\end{aligned}
\]
With $[f]_x$ denoting the coefficient of $x$ in $f\in \mathcal R$, reindexing the coefficient sums gives $[a_i^{s_A}\mathbf X_{ij}]_x=W(a_i^{-s_A}x,i,j)$ and $[\mathbf X_{ij}b_j^{s_B}]_x=W(xb_j^{-s_B},i,j)$.
Thus $\mathbf A\mathbf X=0$ imposes all typed column checks, and $\mathbf X\mathbf B=0$ imposes all typed row checks.
The augmentation $\operatorname{aug}:\mathcal R\to\F$, defined by $\operatorname{aug}(\sum_x c_xx)=\sum_x c_x$, is a ring homomorphism. Applied entrywise, it sends these matrices to $[H_A^0;H_A^1]$ and $[H_B^0;H_B^1]^{\mathsf T}$.
Write $c_X=2r_X$. The full row rank of each stacked matrix allows us to choose pivot port sets of sizes $c_A,c_B$ whose square blocks have invertible augmentations.
The complementary ports are the systematic ports, ordered as $I_A=(I_A(\iota))_{\iota<k_A}$ and $I_B=(I_B(\jmath))_{\jmath<k_B}$.

The input is an arbitrary assignment of $K=k_Ak_Bg$ symbols to the information rectangle $(x,I_A(\iota),I_B(\jmath))$, in the public order~\eqref{eq:public-message-order} of Appendix~\ref{app:public-orders}.
In group-algebra coordinates these are the $k_Ak_B$ entries of $\mathbf X_{I_A,I_B}\in \mathcal R^{k_A\times k_B}$, so $K/N=(k_A/D)(k_B/D)$.
First solve the left system $\mathbf A\mathbf X=0$ on each information column $j\in I_B$ for its pivot row ports. Keep those columns fixed and solve the right system $\mathbf X\mathbf B=0$ on each row $i\in[D]$ for its remaining column ports.

\begin{lemma}[Whole-kernel systematic extension]\label{ltc:systematic}
Every square matrix over $\F[G]$ with invertible augmentation is invertible. Under the stacked rank conditions of Section~\ref{sec:source-hypotheses}, any assignment to the $k_Ak_Bg$ information coordinates above extends uniquely to a codeword of the entire defining-line kernel $\C$.
\end{lemma}
Jennings' theorem~\cite[Theorem~1.2]{Jennings} gives nilpotence of the augmentation ideal, so each pivot is invertible over $\F[G]$. After solving the right system, the left residual $Y=\mathbf A\mathbf X$ vanishes on the fixed columns $I_B$ and satisfies $Y\mathbf B=\mathbf A(\mathbf X\mathbf B)=0$. The invertible right pivot forces the remaining columns of $Y$ to vanish. Both solves are unique, so restriction to the information rectangle is a bijection. The \hyperref[proof:systematic-extension]{proof in Appendix~\ref*{app:systematic-extension}} gives the details.

The finite Neumann series for each pivot inverse can have dense terms.
The encoder avoids forming the inverse by solving the original sparse systems recursively over quotients by central subgroups of order $p$.

For a central subgroup $\langle z\rangle$ of order $p$, the quotient map on $\F[G]$ has kernel $\xi\F[G]$, where $\xi=1-z$ is central.
At each recursion level, the solver runs $p$ rounds. In each round, it solves for the current right-hand side in the quotient and lifts the solution. Subtracting the image of this lift under the system matrix from the current right-hand side gives a residual $r$ whose image in the quotient is zero. The solver chooses the next right-hand side $r'$ to satisfy $\xi r'=r$.
The lifts enter at successive powers of $\xi$. In characteristic $p$, $\xi^p=(1-z)^p=1-z^p=0$, so the residual vanishes after $p$ rounds.
For a group of order $n$, the quotient has order $n/p$ and the nonrecursive work is $O_{D,p}(n)$, giving
\[
 T(n)\le pT(n/p)+O_{D,p}(n)
 \quad\Longrightarrow\quad T(n)=O_{D,p}(n(1+\log n)).
\]
Appendix~\ref{app:central-quotient-solve} constructs the next right-hand side, proves the lifting step for left and right solves, and accounts for the work and space of the $k_B$ left solves and $D$ right solves. These bounds and Lemma~\ref{ltc:systematic} prove Theorem~\ref{thm:encoding}. Section~\ref{sec:fixed-field-family} gives the cost for the fixed-field family.

\subsection{Distance bounds}\label{sec:distance-completion}
Intersection distances force a nonzero codeword to occupy many row or column strips. Typed local distances and port expansion then give the global distance bound.

For a linear code $C$, let $d(C)$ be the minimum Hamming weight of a nonzero codeword, with $d(\{0\})=+\infty$. The normalized $A$- and $B$-port averaging operators send $f\in\mathbb C^G$ to $x\mapsto D^{-1}\sum_i f(a_i^{-1}x)$ and $x\mapsto D^{-1}\sum_j f(xb_j^{-1})$, respectively. Let $\sigma_A,\sigma_B\ge0$ bound their Euclidean operator norms on the mean-zero subspace of $\mathbb C^G$.

\begin{proposition}\label{prop:distance}
Assume the group and rank conditions of the code construction.
For each $X\in\{A,B\}$, let $d_X^{\cap}\ge0$ be an integer satisfying $d(C_X^{\cap})\ge d_X^{\cap}$.
Then
\[
 d(\C)\ge d_A^{\cap}d_B^{\cap}.
\]
Suppose also that integers $d_A,d_B\ge2$ satisfy $d(C_X^s)\ge d_X$ for every $X\in\{A,B\}$ and $s\in\{0,1\}$.
If $0\le\sigma_X<\delta_X:=d_X/D$ for each $X\in\{A,B\}$, then the relative distance $d(\C)/N$ is at least $\Delta$, where
\begin{align}
 \gamma_X&=\frac{\delta_X(\delta_X-\sigma_X)}{1-\sigma_X},&
 \Delta&=\max\left\{\frac{d_A^{\cap}}D\gamma_B,
                    \frac{d_B^{\cap}}D\gamma_A\right\}.\label{ltc:delta}
\end{align}
\end{proposition}

A row strip consists of the coordinates $(x,i,j)$ with $i$ fixed.
The distance of $C_A^{\cap}$ makes any set of fewer than $d_A^{\cap}$ columns of $\operatorname{aug}(\mathbf A)$ independent.
On those columns one can choose a square submatrix of $\mathbf A$ with invertible augmentation, which is invertible over $\F[G]$ by Lemma~\ref{ltc:systematic}.
Thus $\mathbf A\mathbf X=0$ forces every nonzero column of $\mathbf X$ to occupy at least $d_A^{\cap}$ row strips. The symmetric argument for $\mathbf X\mathbf B=0$ gives weight at least $d_B^{\cap}$ in each occupied strip.
For the relative bound, each nonzero row strip is a word of the $B$-port bipartite graph code, whose vertex constraints are the two $B$-typed checks. Each vertex incident to a nonzero edge has at least $d_B$ such edges by local distance. The $B$-port spectral bound then gives strip weight at least $\gamma_BDg$, and interchanging $A$ and $B$ gives the other term in $\Delta$. The \hyperref[proof:distance]{proof in Appendix~\ref*{app:code-distance}} gives the full derivation.

%% file: sections/04-code/completion.tex
\section{Shared Completion and Local Testing}\label{sec:shared-completion}
Shared completion extends consistent legal views to a codeword. It underlies both full-word testing and recovery from partial observations.

\subsection{Tensor lifting and sparse agreement}
Legal rows and columns supplied by neighbouring views prescribe a rectangle at their intersections.
If these values agree and the number of omitted labels in each direction is below the corresponding component-code distance, the rectangle determines a unique legal view preserving the supplied lines~\cite{QuantumTanner,GSW}.

\Needspace{4\baselineskip}
\begin{lemma}\label{lem:tensor-lift}
Let $C_A\subseteq\F^{D_A}$ and $C_B\subseteq\F^{D_B}$ be linear codes.
Let ${\bar I}\subseteq[D_A]$ and ${\bar J}\subseteq[D_B]$ satisfy $|[D_A]\setminus {\bar I}|<d(C_A)$ and $|[D_B]\setminus {\bar J}|<d(C_B)$.
Restriction to ${\bar I}\times {\bar J}$ is a bijection from $C_A\otimes C_B$ onto
\[
 ((C_A|_{\bar I})\otimes\F^{\bar J})\cap(\F^{\bar I}\otimes(C_B|_{\bar J})).
\]
For any rectangle in this image, call its unique preimage its lift.
For $i\in {\bar I}$, any $r\in C_B$ agreeing with row $i$ of the rectangle on ${\bar J}$ equals row $i$ of the lift.
For $j\in {\bar J}$, any $c\in C_A$ agreeing with column $j$ of the rectangle on ${\bar I}$ equals column $j$ of the lift.
\end{lemma}
Puncturing is injective under the deletion bounds, and tensoring the inverse puncturing maps gives the lift (Appendix~\ref{app:tensor-lifting}).

For the four-view construction, fix $d_A,d_B\ge2$ as in Proposition~\ref{prop:distance}, and put $d_{\min}=\min(d_A,d_B)$.

A \emph{checked lift} at type $(s_A,s_B)$ takes omitted row and column sets $I,J\subseteq[D]$ with $|I|+|J|<d_{\min}$, a row in $C_B^{s_B}$ for each $i\in\bar I:=[D]\setminus I$, and a column in $C_A^{s_A}$ for each $j\in\bar J:=[D]\setminus J$. If they agree on $\bar I\times\bar J$, Lemma~\ref{lem:tensor-lift} gives the unique legal view preserving them; otherwise the checked lift fails.

An unassigned view is a \emph{hole}. Its assigned neighbours are its \emph{suppliers} and provide their shared lines.
The following property forces agreement at intersections when the supplier degree bound makes discrepancies sparse.

\begin{definition}[Sparse-agreement property]\label{ltc:je}
Let $e$ be an integer with $1\le e<d_{\min}$.
The local codes satisfy $J_e$ if, for every $s_A,s_B\in\{0,1\}$ and $I,J\subseteq[D]$ with $|I|+|J|\le e$, every matrix in
\[
 (C_A^{s_A}|_{\bar I})\otimes \F^{\bar J}+\F^{\bar I}\otimes(C_B^{s_B}|_{\bar J}),
 \qquad {\bar I}=[D]\setminus I,\quad {\bar J}=[D]\setminus J,
\]
with at most $e-1$ nonzeros in each row and column is zero.
\end{definition}

\subsection{Completing a set of holes}\label{sec:hole-completion}

Shared completion starts with holes $\mathcal Z_0$ and legal matrices $Y_z$ on the other views that agree on shared lines. Partial recovery and the analysis of the line test both construct such assignments.
\begin{enumerate}
\item \emph{Closure.} Set $\mathcal Z=\mathcal Z_0$. Repeatedly unassign a view with at least $e+1$ labelled edges into $\mathcal Z$ and add it to $\mathcal Z$. Return $\bot$ if an addition brings $|\mathcal Z|$ to $2|\mathcal Z_0|$.
\item \emph{Filling.} Set $\mathcal U=\mathcal Z$. While $\mathcal U\ne\varnothing$, choose a view $z\in\mathcal U$ with at most $e$ labelled edges into $\mathcal U$, returning $\bot$ if none exists. Omit the row and column labels of those edges into $\mathcal U$, and apply the checked lift to the lines supplied by assigned neighbours. Return $\bot$ if the lift fails; otherwise assign the lifted view and remove $z$ from $\mathcal U$.
\end{enumerate}
For deterministic recovery, a fixed public rule selects the next view at each step. The guarantee below holds for every allowed sequence of choices.

The number of edges internal to a hole set controls both closure growth and the choice of the next hole to fill.
With $\sigma_A,\sigma_B$ from Section~\ref{sec:distance-completion}, put $\sigma_{\rm avg}=(\sigma_A+\sigma_B)/2$ and $v=4g$.
For a set $\mathcal U$ of $n$ views with internal labelled-edge set $E(\mathcal U)$, the port bounds give
\begin{equation}\label{ltc:mix}
 |E(\mathcal U)|\le D\sigma_{\rm avg}n+
 D(1-\sigma_{\rm avg})\frac{n^2}{v}=: \mu_{\rm mix}(n)
\end{equation}
(see the \hyperref[app:view-edge-bound]{view-graph edge bound in Appendix~\ref*{app:hole-closure}}).

\begin{lemma}[Shared completion]\label{lem:hole-completion}
Assume $J_e$ and $e+1>2D\sigma_{\rm avg}$, and suppose all initially assigned views are legal and agree on shared lines.
Put
\[
 v=4g,\qquad
 \Lambda_e=\frac{v((e+1)/(2D)-\sigma_{\rm avg})}{1-\sigma_{\rm avg}}.
\]
Suppose $2|\mathcal Z_0|<\Lambda_e$.
For every allowed sequence of choices, shared completion succeeds. If $\mathcal Z_0\ne\varnothing$, its closure $\mathcal Z\supseteq\mathcal Z_0$ satisfies
\[
 |\mathcal Z|<2|\mathcal Z_0|,\qquad
 (e+1)(|\mathcal Z|-|\mathcal Z_0|)\le\mu_{\rm mix}(|\mathcal Z|).
\]
If $\mathcal Z_0=\varnothing$, then $\mathcal Z=\varnothing$ and no filling is needed.
Every view outside $\mathcal Z$ has at most $e$ labelled edges into $\mathcal Z$.
The final views are legal and agree on shared lines, defining a codeword $c\in\C$ with $c_z=Y_z$ for every $z\notin\mathcal Z$.
\end{lemma}

Throughout filling, each assigned view has at most $e$ labelled edges into $\mathcal U$. Closure gives this bound initially, and the selection rule gives it for each newly filled view. As $\mathcal U$ shrinks, the bound continues to hold for views already assigned. When filling $z\in\mathcal U$, any retained-crossing conflict forces the opposite view (Figure~\ref{fig:local-square}) into $\mathcal U$. On each supplied line, conflicts use distinct labelled edges into $\mathcal U$ besides the supplier's edge to $z$, leaving at most $e-1$ discrepancies per supplied line. The difference of the row- and column-supplied rectangles lies in the tensor sum of Definition~\ref{ltc:je}; $J_e$ forces agreement, and Lemma~\ref{lem:tensor-lift} fills $z$, preserving its supplied lines.
Appendix~\ref{app:hole-closure} proves the \hyperref[proof:hole-closure]{closure bound} and \hyperref[proof:hole-filling]{filling guarantee}.

The following definition collects the construction, distance, and completion hypotheses used by local testing and partial recovery.

\begin{definition}\label{def:certified-instance}
A \emph{qualified code instance} consists of the construction, public coordinate numbering, and group algorithm data of Section~\ref{sec:source-hypotheses}, the information set of Section~\ref{sec:systematic-encoding}, and bounds $d_X,d_X^{\cap},\sigma_X$ for $X\in\{A,B\}$.
These bounds satisfy the intersection-distance, local-distance, and spectral hypotheses of Proposition~\ref{prop:distance}.
The instance also specifies an integer $1\le e<d_{\min}$ for which property $J_e$ holds and
\begin{equation}\label{eq:qualified-closure-margin}
 e+1>2D\sigma_{\rm avg}.
\end{equation}
\end{definition}

\subsection{Local testability}
\label{sec:testing}

Shared completion also relates the line test to distance from the code. To test a full word $W$, draw a labelled line uniformly and check its parity equations. This gives rejection probability $\beta(W)=\ill(W)/M$.

\begin{theorem}\label{thm:testing}
Every qualified code instance admits positive constants $\beta_0,L_{\rm r}$, depending only on its local and spectral data, with the following guarantees.
The line test has perfect completeness and reads $D$ field symbols.
For every $W\in\F^N$,
\[
 \beta(W)\ge\frac{\beta_0}{1-K/N}\dist(W,\C).
\]
If $\beta(W)\le\beta_0$, there exists $c\in\C$ with $\dist(W,c)\le L_{\rm r}\beta(W)$.
\end{theorem}

For $\beta(W)\le\beta_0$, apply Lemma~\ref{lem:tensor-lift} to legal lines in views with fewer than $d_{\min}$ illegal labels in each direction, marking others as forced holes. Unassign a minimum vertex cover of the bipartite conflict graph on the lifted views so the remaining views agree on shared lines. The forced holes and the cover form the initial hole set $\mathcal Z_0$. By K\"onig's theorem, the cover has the size of a maximum matching. The matched edges have disjoint endpoints among the lifted views. Each matched pair and each forced hole accounts for at least $d_{\min}$ illegal incidences. Since each illegal line has two incidences, $|\mathcal Z_0|\le2\ill(W)/d_{\min}$. Completion yields $c\in\C$, with $\dist(W,c)$ controlled by lift changes outside and entries inside the closure. For $\beta(W)>\beta_0$, systematic encoding gives $\dist(W,\C)\le 1-K/N$, as required in this case.
The \hyperref[proof:local-testability]{proof in Appendix~\ref*{app:testing-proof}} gives the seed construction, distance count, and constants.

%% file: sections/05-recovery/recovery.tex
\section{Recovery from Partial Observations}\label{sec:recovery}
Shared completion gives the distance bound for full-word testing. The DAS extractor uses the partial-recovery guarantee below: recovery first builds legal views that agree on shared lines from the observations $y\in\F^S$. The initialization defect count $\ill_{\rm init}(y)$ is computed directly from $y$.
Under the local-consistency condition below, $\ill_{\rm init}(y)\le\ill(W)$ for every full word $W$ with $W|_S=y$ (Lemma~\ref{lem:recovery-initialization}).
The defect and erasure counts bound the initial holes for shared completion under admissible parameters.

\subsection{Recovery algorithm}\label{sec:recovery-algorithm}

\paragraph{Public configuration and local observations.}
Fix a qualified instance, a positive integer distance lower bound $d_C\le d(\C)$, and public nonnegative integer parameters $(h,\ill_{\max},R,u,t)$ with $\ill_{\max}<M$ and $u<N$.
Initialization uses a public integer threshold $\kinit\ge1$, and $e$ governs closure and filling.

A view is \emph{heavy} if it is incident to an erasure-heavy line.
A non-heavy line is \emph{incompatible} if no legal line word matches all its observations. Such a line is illegal in every full word extending $y$.
The local-consistency condition $2(\kinit-1)+h<d_{\min}$ implies $h<d_{\min}$, so each non-heavy line has at most one legal completion. It also gives $\kinit<d_{\min}$, as required for the initialization lifts.

\paragraph{Partial line completion.}
Let $\mathcal L$ have ordered physical coordinates $(\mathsf{pos}_j)_{j\in[D]}$ and local parity-check matrix $H_{\mathcal L}$, with columns in that order.
For $U_{\mathcal L}=\{j\in[D]:\mathsf{pos}_j\notin S\}$ with $|U_{\mathcal L}|\le h<d_{\min}$, the missing columns of $H_{\mathcal L}$ are independent. Writing $H_{\mathcal L}|_T$ for the submatrix of columns indexed by $T\subseteq[D]$, a legal completion is the unique solution $\eta_{\mathcal L}\in\F^{U_{\mathcal L}}$, if one exists, to
\begin{equation}
(H_{\mathcal L}|_{U_{\mathcal L}})\,\eta_{\mathcal L}
=-(H_{\mathcal L}|_{[D]\setminus U_{\mathcal L}})\,
  (y(\mathsf{pos}_j))_{j\in[D]\setminus U_{\mathcal L}}.
\label{eq:partial-line}
\end{equation}

Figure~\ref{fig:recovery-algorithm} gives the two stages of recovery, including the failure checks.

\begin{figure}[!htb]
\centering
\setlength{\fboxsep}{6pt}
\fbox{\begin{minipage}{\dimexpr\linewidth-2\fboxsep-2\fboxrule\relax}
\small
\textbf{Input.} A partial word $y\in\F^S$, with observed coordinates $S\subseteq[N]$.
\par\textbf{Output.} A codeword $c\in\C$ or $\bot$. The budgets $u$ and $t$ bound total erasures and disagreements on $S$, respectively.
\par\smallskip
\textbf{Stage I. Line completion and initialization.}
\begin{enumerate}
\item \emph{Line completion.} Mark a line as heavy if it has more than $h$ unobserved coordinates. On each non-heavy line solve~\eqref{eq:partial-line}. Store the unique solution when it exists and mark the line incompatible otherwise.
\item \emph{Initialization.} Leave a view unassigned if it is incident to a heavy line or has more than $\kinit$ incompatible incident labels. At every other view, omit its incompatible row and column labels and apply the checked lift to its retained completed lines. A successful lift assigns a legal view, and a failed lift leaves the view unassigned. Write $\mathcal Z_0$ for the resulting set of initial holes.
\end{enumerate}

\textbf{Stage II. Shared completion and verification.}
\begin{enumerate}
\setcounter{enumi}{2}
\item \emph{Closure.} Run the closure stage of Section~\ref{sec:hole-completion} on $\mathcal Z_0$.
\item \emph{Filling.} Run the filling stage on the resulting hole set. Both stages use the fixed public choice rules and return $\bot$ if they fail.
\item \emph{Merge and verify.} Merge the assigned views in fixed order, returning $\bot$ on any coordinate conflict. For the merged word $c$, explicitly verify $c\in\C$, $N-|S|\le u$, and $\wt(c|_S-y)\le t$. Return $c$ if all three checks pass, and $\bot$ otherwise.
\end{enumerate}
\end{minipage}}
\caption{Deterministic checked recovery $\mathsf{Rec}$ on the partial word $y\in\F^S$.}\label{fig:recovery-algorithm}
\end{figure}

\paragraph{Observable initialization count.}
For each full extension $W$ of $y$, incompatible lines are illegal in $W$, and each non-heavy initial hole meets at least $\kinit+1$ lines illegal in that $W$ under the local-consistency condition.
Let $\inc(y)$ count incompatible non-heavy lines and $\inc_z$ count such lines incident to an assigned view $z$.
On side $i\in\{0,1\}$, let $n_i$ count non-heavy initial holes.
Each labelled line has one endpoint per side, so distinct views on the same side have disjoint sets of incident lines.
We charge $\kinit+1$ to each non-heavy initial hole and $\inc_z$ to each assigned view.
The global count $\inc(y)$ also includes incompatible non-heavy lines joining two heavy views.
Define
\begin{equation}\label{eq:initialization-budget}
\ill_{\rm init}(y)=\max\Biggl\{\inc(y),\,\max_{i\in\{0,1\}}\Biggl[(\kinit+1)n_i+
\sum_{\substack{z\text{ assigned}\\\text{on side }i}}\inc_z\Biggr]\Biggr\}.
\end{equation}
Under local consistency, each count, and hence their maximum, is at most $\ill(W)$ for every full extension $W$ (Lemma~\ref{lem:recovery-initialization}).
\begin{lemma}[Initialization comparison]\label{lem:recovery-initialization}
Assume $2(\kinit-1)+h<d_{\min}$. The views assigned at initialization are legal, preserve all retained completed lines, agree along lines shared by two assigned views, and can differ from their observations only at crossings of incompatible rows and columns. For every $S\subseteq[N]$, $y\in\F^S$, and every full word $W$ with $W|_S=y$, each non-heavy initial hole is incident to at least $\kinit+1$ lines illegal in $W$, and
\[
 \ill_{\rm init}(y)\le\ill(W).
\]
If also $\kinit\le e$ and $h\le e-1$, then $\ill_{\rm init}(y)=\inc(y)$ for every partial input $y$.
\end{lemma}

The \hyperref[proof:initialization-comparison]{proof is in Appendix~\ref*{app:recovery-initialization-proof}}.

\Needspace{6\baselineskip}
If the observations come from a codeword and a row is the only heavy line, Stage~I leaves only its two endpoint views unassigned.
If at most $e$ labelled edges join any pair of views, closure adds no holes because an outside view can meet at most one of the two opposite-side endpoints.
Filling lifts either endpoint with their shared rows omitted (including any parallel rows), then lifts the other with all lines supplied.

\subsection{From initialization to admissible budgets}\label{sec:recovery-parameters}
Assume $\ill_{\rm init}(y)\le\ill_{\max}$, $r_{\rm H}(S)\le R$, and $N-|S|\le u$.
Let $H_0$ and $I_0$ count heavy and non-heavy initial holes.
Each heavy line marks one endpoint per side, and each heavy hole contains at least $h+1$ erased corners.
There are at most $2u$ erased corners per side because each missing coordinate has two corners there (Figure~\ref{fig:local-square}).
Each non-heavy hole uses $\kinit+1$ of its side charge in~\eqref{eq:initialization-budget}.
Thus $H_0\le 2L_*$, $I_0\le 2q_*$, and $H_0+I_0\le n_{\rm hole}$, where
\begin{equation}\label{eq:recovery-hole-budgets}
 \begin{aligned}
 L_*&=\min\{R,\lfloor2u/(h+1)\rfloor\},&
 q_*&=\lfloor\ill_{\max}/(\kinit+1)\rfloor,\\
 n_{\rm hole}&=2(L_*+q_*).
 \end{aligned}
\end{equation}

\paragraph{Observed disagreements.}
To count disagreements inside the closure, first bound its size. Set $\mathsf C_e(0)=0$.
For an integer $0<s<\Lambda_e/2$, let $\mathsf C_e(s)$ be the largest integer $n\in[s,2s)$ with $(e+1)(n-s)\le\mu_{\rm mix}(n)$.
Here $\Lambda_e$ is the closure capacity from Lemma~\ref{lem:hole-completion}, and $\mu_{\rm mix}$ is the edge bound in~\eqref{ltc:mix}.
Then $\mathsf C_e(s)$ bounds the closure size for any seed of at most $s$ holes (see the \hyperref[app:integer-closure-bound]{integer closure bound in Appendix~\ref*{app:recovery-budgets}}).

Each disagreeing coordinate contributes four corners.
The closure contains at most $D^2\mathsf C_e(H_0+I_0)$ corners, of which at least $(h+1)H_0$ are erased and cannot disagree with $y$.
Outside the closure, disagreements can occur only at crossings of incompatible rows and columns (Lemma~\ref{lem:recovery-initialization}). One upper bound counts distinct row--column pairs. The other combines a per-view crossing bound with the side charges left after accounting for non-heavy holes.
The resulting upper bound on disagreeing corners is
\begin{equation}\label{eq:recovery-corner-bound}
\begin{aligned}
 F(H_0,I_0)&=D^2\mathsf C_e(H_0+I_0)-(h+1)H_0\\
           &\quad+\min\{\ill_{\max}^2/4,\ \kinit[2\ill_{\max}-(\kinit+1)I_0]/4\}.
\end{aligned}
\end{equation}
\begin{lemma}[Observed disagreement bound]\label{lem:residual-budget}
Fix a qualified instance and observations $y\in\F^S$ with $\ill_{\rm init}(y)\le\ill_{\max}$, $r_{\rm H}(S)\le R$, and $N-|S|\le u$.
Assume $2(\kinit-1)+h<d_{\min}$ and $2n_{\rm hole}<\Lambda_e$.
Then Steps~3--4 of Figure~\ref{fig:recovery-algorithm} complete the initialized views as in Lemma~\ref{lem:hole-completion}, and the codeword $c\in\C$ given by this shared completion satisfies
\begin{equation}\label{eq:recovery-residual}
 \wt(c|_S-y)\le\widehat{\mathsf T}=\left\lfloor F(2L_*,2q_*)/4\right\rfloor .
\end{equation}
\end{lemma}
The \hyperref[proof:observed-disagreements]{proof in Appendix~\ref*{app:recovery-budgets}} derives the crossing counts and shows that $F$ is nondecreasing in both hole counts up to $2L_*$ and $2q_*$, respectively.
To combine these bounds with the distance requirement, we call the recovery parameters \emph{admissible} if
\begin{equation}\label{eq:admissible-basic}
\begin{aligned}
 &2(\kinit-1)+h<d_{\min},\qquad 2n_{\rm hole}<\Lambda_e,\\
 &\widehat{\mathsf T}\le t
 \le\left\lfloor\frac{d_C-1}{2}\right\rfloor-u.
\end{aligned}
\end{equation}
The lower bound on $t$ makes the disagreement check pass, and the upper bound gives $2(t+u)<d_C$ for uniqueness.
Section~\ref{sec:family-guarantee} uses a simpler upper bound on $\widehat{\mathsf T}$ to choose the fixed-family budgets.
Section~\ref{sec:finite-instance} gives a concrete set of admissible parameters satisfying~\eqref{eq:admissible-basic}.

\subsection{Recovery guarantees}
\label{sec:recovery-guarantees}

\begin{theorem}\label{thm:recovery}
Fix a qualified code instance and admissible parameters satisfying~\eqref{eq:admissible-basic}.
The recovery algorithm $\mathsf{Rec}$ is deterministic and has the following guarantees.

For every $S\subseteq[N]$ and $y\in\F^S$, recovery succeeds if
\[
 \ill_{\rm init}(y)\le\ill_{\max},\qquad
 r_{\rm H}(S)\le R,\qquad N-|S|\le u.
\]
If $\kinit\le e$ and $h\le e-1$, $\inc(y)\le\ill_{\max}$ may replace \mbox{$\ill_{\rm init}(y)\le\ill_{\max}$}.

In particular, for every $W\in\F^N$ and $S\subseteq[N]$, recovery succeeds on $W|_S$ whenever
\[
 \ill(W)\le\ill_{\max},\qquad
 r_{\rm H}(S)\le R,\qquad N-|S|\le u.
\]

Moreover, whenever recovery returns a word $c$ on any $y\in\F^S$, it satisfies
\[
 c\in\C,\qquad N-|S|\le u,\qquad \wt(c|_S-y)\le t.
\]
\end{theorem}

\begin{proof}
Fix $S\subseteq[N]$ and $y\in\F^S$ satisfying the three input budgets.
Because $2(\kinit-1)+h<d_{\min}$ by~\eqref{eq:admissible-basic}, Stage~I produces legal, agreeing assigned views (Lemma~\ref{lem:recovery-initialization}).
The input budgets and admissibility~\eqref{eq:admissible-basic} satisfy the hypotheses of Lemma~\ref{lem:residual-budget}, so Steps~3--4 succeed and their shared completion merges into a codeword $c\in\C$ satisfying~\eqref{eq:recovery-residual}.
Since $\wt(c|_S-y)\le\widehat{\mathsf T}\le t$ by~\eqref{eq:admissible-basic} and $N-|S|\le u$, all three final checks in Figure~\ref{fig:recovery-algorithm} pass and recovery returns $c$.

When $\kinit\le e$ and $h\le e-1$, Lemma~\ref{lem:recovery-initialization} gives $\ill_{\rm init}(y)=\inc(y)$, proving the alternative criterion.
For every full word $W\in\F^N$ and every $S\subseteq[N]$, the same lemma gives $\ill_{\rm init}(W|_S)\le\ill(W)$, so applying the first guarantee to $y=W|_S$ proves the DAS recovery statement.
On arbitrary inputs, Step~5 returns a word only after explicitly verifying code membership, the erasure bound, and observed disagreements.
\end{proof}

\begin{corollary}\label{cor:instance-interface}
Fix a qualified code instance, a positive integer distance lower bound $d_C\le d(\C)$, and admissible parameters $(\kinit,h,\ill_{\max},R,u,t)$.
With the integers $(h,\ill_{\max},R,u,t)$ and the recovery algorithm of Section~\ref{sec:recovery-algorithm} as $\mathsf{Rec}$, the code is a line-recovery code (Definition~\ref{def:line-recovery}).
\end{corollary}

Appendix~\ref{app:instance-interface-proof} verifies the interface conditions.

\paragraph{Work.}
After reusable preparation, recovery uses $O_{\rm fam}(N)$ code-field and index operations. Proposition~\ref{prop:recovery-cost} gives the work and workspace bounds for a first-in, first-out queue implementation.
Appendix~\ref{app:full-resource-ledger} accounts for preparation, authentication, and transcript handling.

%% file: sections/06-das/security.tex
\section{Fixed-Field Instantiation and One-Line DAS}
\label{sec:security}

We construct a uniform qualified family over the fixed code field and choose admissible budgets to instantiate Theorem~\ref{thm:das} with one line per client. We give resource bounds and a concrete sampling configuration.

\input{sections/04-code/uniform-family}

\input{sections/07-comparison/parameters}

%% file: sections/04-code/uniform-family.tex
\subsection{A uniform fixed-field family}\label{sec:fixed-field-family}
\label{sec:local-certificate}\label{sec:local-preprocessing}
We first obtain local checks for a fixed code field and target rate.
An arithmetic tower will then let us increase the block length while keeping these checks and the line length fixed.

\begin{lemma}[Qualified local checks]
\leavevmode\vadjust pre{\hbox{\phantomsection\label{lem:qualified-local-checks}}}%
Fix $\F=\mathbb F_q$ and a target rate $\rho\in(0,1)$, and put $\bar r=(1-\sqrt\rho)/2$.
There are constants $\bar e>0$ and $c>0$ such that, for all sufficiently large $D$, the integers
\[
 r=\lfloor\bar rD\rfloor,\qquad e=\lfloor\bar eD\rfloor,\qquad
 d=d_{\cap}=\lfloor4\bar eD\rfloor
\]
satisfy $2r<D$, $2\le e\le r$, and $e<d\le D$.
Draw two independent uniform full-row-rank $2r\times D$ stacks and split each into two $r$-row checks.
Except with probability at most $q^{-cD}$, all typed distances are at least $d$, both intersection distances are at least $d_{\cap}$, and all four typed pairs satisfy $J_e$.
In particular, such fixed checks exist, and $(1-2r/D)^2\ge\rho$.
\end{lemma}

Appendices~\ref{app:fullrank-asymptotic}--\ref{app:fixed-field-checks} prove the qualification bound and this parameter choice. The checks can be selected once by exhaustive search at a cost depending only on $(D,q)$ and then reused along the tower.

\phantomsection\label{sec:groups-family}
Congruence quotients at a ramified place of a quaternion division algebra supply the central tower for encoding. At a split place, the Ramanujan spectrum of the associated Bruhat--Tits tree quotients gives the uniform port bounds used for distance and shared completion.
We take the arithmetic field $\mathbb F_Q$ in the same characteristic as the fixed code field $\F$ and set $D=Q+1$.

\begin{lemma}[Arithmetic $p$-group tower]\label{lem:arithmetic-spectrum}
For every prime $p$ and integer $a\ge1$, put $Q=p^a$ and $D=Q+1$.
There is a tower of finite $p$-groups $(G_m)_{m\ge0}$ with $|G_m|=p^m$, each equipped with $D$ labelled ports.
In each $G_m$, the same labelled list supplies the left ports $a_i$ and the right ports $b_i$.
This list includes the identity, and distinct labels may give the same element of $G_m$.
Each map $G_{m+1}\twoheadrightarrow G_m$ has central kernel of order $p$ and sends every port to the port with the same label.
For every $m\ge0$, the normalized $A$- and $B$-port averaging operators on the mean-zero subspace of $\mathbb C^{G_m}$ have Euclidean operator norms at most $\sigma_D=2\sqrt{D-1}/D=2\sqrt Q/(Q+1)$, so one may take $\sigma_A=\sigma_B=\sigma_D$.
For fixed $D$, constructing $G_m$ with the quotient-chain data read by the encoder (Appendix~\ref{app:central-quotient-solve}) and the public group index $\operatorname{idx}$ costs $T_G=O_D(g(1+\log g))$ arithmetic-field operations over $\mathbb F_Q$, together with array and index operations, where $g=|G_m|$.
\end{lemma}

The \hyperref[proof:arithmetic-spectrum]{proof in Appendix~\ref*{app:family-assembly}} combines the labelled-port spectral transfer (Lemma~\ref{lem:port-spectral-transfer}) with the effective central tower and numbering (Lemma~\ref{lem:effective-p-tower}).

To combine the local checks with the tower, choose $D=p^a+1$ large enough for Lemma~\ref{lem:qualified-local-checks} and the inequalities $d/D>\sigma_D$ and $e+1>2D\sigma_D$. Such a choice exists because $d/D\to4\bar e>0$, $e/D\to\bar e>0$, and $\sigma_D\to0$. With these checks and $D$ fixed, increasing $m$ gives the following family.

\begin{proposition}[Uniform fixed-field family]\label{prop:family}
For every finite field $\F=\mathbb F_q$ of characteristic $p$ and every target rate $\rho\in(0,1)$, there exist fixed local parameters and local checks, and an unbounded family $\mathcal F=(\C_m)_{m\ge0}$ over $\F$, with the following properties.
\begin{enumerate}
\item \emph{Length, rate, and qualification.}
The member indexed by $m$ has length $N=D^2p^m$ and rate $K/N\ge\rho$.
Each member is a qualified code instance (Definition~\ref{def:certified-instance}), and all members have a common positive relative-distance lower bound $\Delta$.

\item \emph{Testing and completion.}
Every member has the $D$-query tester and distance bounds of Theorem~\ref{thm:testing}, with common positive constants $\beta_0,L_{\rm r}$.
The normalized closure capacity $\Lambda_e/v$ is positive and independent of $m$.

\item \emph{Systematic encoding.}
The encoder is a systematic bijection onto the line-defined code. After group preparation, encoding uses $O_{\rm fam}(N\log N)$ code-field and index operations by Theorem~\ref{thm:encoding}; the same asymptotic bound applies separately to the arithmetic-field and index operations for reusable group preparation $T_G$.
\end{enumerate}
The code and arithmetic fields, $D$, the local checks, and their distance, spectral, and sparse-agreement bounds remain fixed as $m$ grows.
\end{proposition}

Appendix~\ref{app:family-assembly} proves this fixed-family construction and its target-rate choice.

%% file: sections/07-comparison/parameters.tex
\input{sections/07-comparison/asymptotic-comparison}

\subsection{Resource bounds and comparison}\label{sec:parameters}\label{sec:family-resources}
With the sampling threshold and digest width fixed, we account for the resulting communication and computation.

\paragraph{Communication.}
Table~\ref{tab:sampling-costs} compares asymptotic publisher--client communication for Hash, FRIDA, and our scheme (for a fixed family with $\log\Qhash=O(\lambda)$). Since honest publication takes $2N$ hashes, this cap requires $\log N=O(\lambda)$.
Under this cap, setting $\kappa$ to the threshold in~\eqref{eq:fixed-das-digest-width} gives $\Theta(\lambda)$-bit digests and an $O_{\rm fam}(\lambda\log K)$-bit first-hop exchange per draw.

Appendix~\ref{app:cross-scheme-comparison} derives our first-hop bound.

\begin{table}[htb]
\caption{Asymptotic first-hop publisher--client communication bounds (big-$O$ omitted) for $K$ information symbols and security parameter $\lambda$, at constant rate and symbol width. Total draws across clients suffice for reconstruction under each scheme's stated conditions. A draw requests a column in Hash~\cite[Section~9.1]{Foundations}, a position in FRIDA, or a $D$-symbol line here. The Hash and FRIDA bounds are from~\cite[Section~5, Table~1]{FRIDA}. FRIDA assumes input sizes supported by its field and fixed FRI fan-in and batch size. For our fixed family, $N/D=\Theta_{\rm fam}(K)$. The hidden constants may depend on the family, including $D$. Bounds involving the draw count may also depend on the erasure fraction $\zeta$ (Appendices~\ref{app:asymptotic-details} and~\ref{app:cross-scheme-comparison}). Our bounds apply to sufficiently large members when the complete experiment has $\log\Qhash=O(\lambda)$ (hence $\log N=O(\lambda)$). Appendix~\ref{app:full-resource-ledger} gives our costs for context delivery, reusable public data, and transcript forwarding.}

\label{tab:sampling-costs}
\centering\small
\setlength{\tabcolsep}{4pt}
\begin{tabular}{@{}lccc@{}}
\toprule
Quantity & Hash & FRIDA & This work \\
\midrule
Commitment (bits) & $\lambda\sqrt{K}$ & $\lambda^2\log^2 K$ & $\lambda$ \\
Exchange per draw (bits) & $\sqrt{K}$ & $\lambda\log^2 K$ & $\lambda\log K$ \\
Sufficient total draws & $\sqrt{K}+\lambda$ & $K+\lambda$ & $K+\lambda$ \\
Sampling exchange (bits) & $(\sqrt{K}+\lambda)\sqrt{K}$ & $(K+\lambda)\lambda\log^2 K$ & $(K+\lambda)\lambda\log K$ \\
\bottomrule
\end{tabular}
\end{table}

\FloatBarrier
\paragraph{Work.}
With fixed local checks, preparing the group and addressing data takes $O_{\rm fam}(N\log N)$ arithmetic-field and index operations.
Each publication then costs $O_{\rm fam}(N\log N)$ code-field and index operations and $2N$ hashes.
With compact client addressing data, each one-line client uses $O_{\rm fam}((1+\log N)^3)$ arithmetic-field and index operations to compute the requested addresses (Lemma~\ref{lem:public-addressing}), $O_{\rm fam}(1)$ code-field operations for the line check, and $O_{\rm fam}(\log N)$ Merkle hashes. The \hyperref[app:public-data-charges]{public-data charges in Appendix~\ref*{app:full-resource-ledger}} include delivery or local generation of the addressing data.
With compiled addressing tables, the recovery party uses $O_{\rm fam}(N+n_{\rm dr})$ code-field and index operations for addressing, merging, and algebraic recovery.
The extractor uses $O_{\rm fam}(n_{\rm dr}\log N)$ Merkle hashes to verify a complete list (one of the two verification passes in~\eqref{eq:all-authentication-hashes}).

\Needspace{10\baselineskip}
\input{sections/06-das/finite-instance}

%% file: sections/07-comparison/asymptotic-comparison.tex
\subsection{Proof of the one-line DAS theorem}
\label{sec:family-guarantee}

To prove Theorem~\ref{thm:main-das} for a given finite field $\F$ and target rate $\rho\in(0,1)$, let $\mathcal F$ be the qualified family supplied by Proposition~\ref{prop:family}, which has rate $K/N\ge\rho$.
All members of $\mathcal F$ share the code field, line length $D$, local checks, distance and spectral bounds, and sparse-agreement parameter $e$, as well as a positive relative-distance lower bound $\Delta$.
Along the tower, $N=D^2g$, $M=4Dg$, and $K=\Theta_{\rm fam}(N)$. Put $d_C=\lceil N\Delta\rceil$. The positive normalized closure capacity $\Lambda_e/v$ is independent of the member index $m$.

\begin{lemma}[Fixed-family budgets and draw count]
\label{lem:family-budgets}
Fix $0<\zeta<\Delta/2$. Above a member-size threshold depending only on $\mathcal F$ and $\zeta$, every pair of integers $\kinit\ge1$ and $h\ge0$ with $2(\kinit-1)+h<d_{\min}$ has admissible budgets as in~\eqref{eq:admissible-basic}, with
\[
 \ill_{\max}=\Theta_{\rm fam,\zeta}(M),\qquad
 R=\Theta_{\rm fam,\zeta}(g),\qquad
 t=\Theta_{\rm fam,\zeta}(N),\qquad
 u=\lfloor\zeta N\rfloor.
\]
Both the recovery budgets and the member-size threshold are chosen independently of $\lambda$ and apply for every integer $\lambda\ge1$.

There is a constant $c_{\rm draw}>0$, depending only on $\mathcal F$ and $\zeta$, such that for those budgets and every integer $\lambda\ge1$, some integer $n_{\rm req}\le c_{\rm draw}(N/D+\lambda)$ has the following properties.
The avoidance bound~\eqref{eq:lt-prob} is at most $2^{-(\lambda+3)}$ for $n_{\rm dr}\ge n_{\rm req}$.
For every registration history and supported set of selected clients with $n_{\rm dr}\ge n_{\rm req}$ independent uniform draws, the requested coordinates satisfy~\eqref{eq:observation-failure} with $\eps_{\rm obs}=2^{-(\lambda+3)}$.
\end{lemma}

\begin{proof}
Set $C_{\rm res}=M/[g(\kinit+1)]=4D/(\kinit+1)$.
Since $\Lambda_e/v>0$ and $2\zeta<\Delta$, choose fixed $\theta,\eta\in(0,1)$ with
\begin{equation}\label{eq:family-admissible-fractions}
 \eta+C_{\rm res}\theta<\Lambda_e/v,\qquad 2(C_{\rm res}\theta+\eta+\zeta)<\Delta.
\end{equation}
The first margin permits closure. The second requires $C_{\rm res}\theta+\eta<\Delta/2-\zeta$, leaving room to round the budgets.
Use the integer budgets
\begin{equation}\label{eq:family-positive-budgets}
 \ill_{\max}=\lfloor\theta M\rfloor,\quad R=\lfloor\eta g\rfloor,\quad
 u=\lfloor\zeta N\rfloor,\quad t=\lceil N(C_{\rm res}\theta+\eta)\rceil.
\end{equation}
The caps $L_*\le R\le\eta g$ and $q_*\le\ill_{\max}/(\kinit+1)\le C_{\rm res}\theta g$ give $2n_{\rm hole}/v=(L_*+q_*)/g\le\eta+C_{\rm res}\theta<\Lambda_e/v$.
The \hyperref[app:fixed-residual-estimate]{residual estimate in Appendix~\ref*{app:fixed-budget-check}} gives
\[
 \widehat{\mathsf T}\le D^2\left(L_*+\frac{\ill_{\max}}{\kinit+1}\right)
 \le N(C_{\rm res}\theta+\eta)\le t.
\]
For sufficiently large $g$, rounding preserves the stated budget orders and the second margin in~\eqref{eq:family-admissible-fractions} gives $2(t+u)<d_C$, establishing admissibility~\eqref{eq:admissible-basic}.

The cap on $V$ enforces the total erasure budget, while the cap on $r_{\rm H}(S^*)$ limits the initial holes caused by heavy lines.
Appendix~\ref{app:asymptotic-details} combines the illegal-line avoidance bound~\eqref{eq:lt-prob} with the fixed-draw bounds for $V$ and $r_{\rm H}(S^*)$ in Appendix~\ref{app:occupancy}.
The resulting threshold~\eqref{eq:family-line-only-counts} meets both error bounds with $n_{\rm req}\le c_{\rm draw}(N/D+\lambda)$.
Only finitely many $(\kinit,h)$ satisfy the local-consistency condition, so the threshold and draw constant can be chosen uniformly.
\end{proof}

To apply the generic security theorem (Theorem~\ref{thm:das}) at security parameter $\lambda$, fix both the public resource bounds and the Merkle digest width $\kappa$ before message choice as follows.
The cap $\Qhash$ on all oracle queries in the complete experiment is fixed independently of $\kappa$. The resource bounds must allow a complete honest run with at least $n_{\rm req}$ clients (see~\eqref{eq:all-authentication-hashes} and the query accounting in Appendix~\ref{app:full-resource-ledger}).
Choose $\kappa$ to satisfy
\begin{equation}\label{eq:fixed-das-digest-width}
 \kappa\ge\lambda+2+\left\lceil\log_2(3\Qhash^2+\Qhash)\right\rceil.
\end{equation}

\begin{proof}[Theorem~\ref{thm:main-das}]
\phantomsection\label{proof:main-das-composition}
For the family of Proposition~\ref{prop:family}, fix a locally consistent pair $(\kinit,h)$ and take its budgets and $n_{\rm req}$ from Lemma~\ref{lem:family-budgets}.
Set $a_{\rm cl}=1$ and $\ell_0=n_{\rm req}$, and use the Merkle instance of Appendix~\ref{app:authentication-instance} with $\kappa$ satisfying~\eqref{eq:fixed-das-digest-width}.

For completeness and soundness, the selected set has $\ell\ge n_{\rm req}$ clients and makes $n_{\rm dr}=\ell$ independent draws after registration. Lemma~\ref{lem:family-budgets} bounds the avoidance and observation errors. Apply Theorem~\ref{thm:das} with Corollary~\ref{cor:instance-interface} and Lemma~\ref{lem:auth-instance}.
Position binding gives consistency for any two supported lists under the same registration. The three errors satisfy
\[
\begin{gathered}
\eps_{\rm comp}(\ell)\le2^{-(\lambda+3)},\qquad
\eps_{\rm snd}(\ell)\le2^{-(\lambda+3)}+\min\{1,(3\Qhash^2+\Qhash)/2^\kappa\},\\
\eps_{\rm cons}\le\min\{1,\Qhash(\Qhash-1)/2^{\kappa+1}\}.
\end{gathered}
\]
By~\eqref{eq:fixed-das-digest-width}, each authentication term is at most $2^{-(\lambda+2)}$.
Since $2^{-(\lambda+3)}+2^{-(\lambda+2)}<2^{-\lambda}$, all three errors are at most $2^{-\lambda}$.
\end{proof}

%% file: sections/06-das/finite-instance.tex
\subsection{A concrete one-line DAS configuration}\label{sec:finite-instance}
Take $G_1$ from Lemma~\ref{lem:arithmetic-spectrum} at $p=2$, $D=8193$, and choose local checks over $\F_{2^{64}}$ by Lemma~\ref{lem:finite-local-qualification} (Appendix~\ref{app:fullrank-asymptotic}). Here $g=2$, $N=134{,}250{,}498$, and $M=65{,}544$. The code carries about $269$ MB of information at rate $(4097/8193)^2>1/4$. This finite configuration illustrates how the general bounds certify the one-line protocol (Appendix~\ref{app:finite-recovery-parameters}).
\begin{center}
\small
\begin{tabular}{@{}P{.27\linewidth}@{\quad}P{.67\linewidth}@{}}
\toprule
Local checks & $r=2048$, $e=384$, $d=1900$, $d_{\cap}=3900$ \\
Recovery & $\kinit=\ill_{\max}=105$, $h=1690$, $R=0$, $u=1{,}400{,}000$, $t=689$ \\
128-bit security & $100{,}000$ draws, $256$-bit digests, $\Qhash=2^{62}$ \\
\bottomrule
\end{tabular}
\end{center}
At $g=2$ the closure condition $2n_{\rm hole}<\Lambda_e<2$ requires $n_{\rm hole}=0$. With these budgets ($u/N\approx1.04\%$), Stage~I assigns all eight views. The same algorithm fills holes for the positive budgets of Section~\ref{sec:family-guarantee}. Each client receives $8193$ field symbols and their Merkle multiproof. The four reusable local check matrices take about $537$ MB in dense form, and each line check uses at most $16{,}779{,}264$ field multiplications.

%% file: sections/supplement-guide.tex
\section*{Supplementary Material}
\phantomsection\label{supplement-start}
\begin{center}
\small
\begin{tabular}{@{}P{.36\linewidth}@{\quad}P{.60\linewidth}@{}}
\toprule
Main-text result or use & Proof or supporting material \\
\midrule
Authentication (\S\ref{sec:generic-security}) & Appendix~\ref{app:authentication-instance}: tree format, reference word $W$ and authentication bounds (Lemma~\ref{lem:auth-instance}), and multiproof sizes \eqref{eq:multiproof-identity} and~\eqref{eq:multiproof-frontier} \\
Public orders, encoding, and distance (\S\ref{sec:ltc}) & Appendix~\ref{ltc:app-count}: coordinate and message orders, systematic extension, central-quotient solves, and Proposition~\ref{prop:distance} \\
Shared completion, testing, and recovery (\S\S\ref{sec:shared-completion},~\ref{sec:recovery}) & Appendix~\ref{app:completion-testing}: tensor lifting, closure and filling, full-word testing, and the recovery guarantees \\
Qualified fixed-field family (\S\ref{sec:fixed-field-family}) & Appendix~\ref{app:arithmetic-family}: \hyperref[app:fullrank-asymptotic]{local qualification}, \hyperref[app:arithmetic-setup]{the arithmetic tower}, \hyperref[app:family-assembly]{family assembly}, and \hyperref[app:finite-recovery-parameters]{finite-instance parameters} \\
Addressing and recovery cost (\S\S\ref{sec:recovery},~\ref{sec:family-resources}) & Appendix~\ref{app:addressing}: compact data, recovery tables, reordering, and recovery work and space \\
Budgets, sampling, and resources (\S\S\ref{sec:family-guarantee},~\ref{sec:parameters}) & Appendix~\ref{app:collective-das}: fixed-family budget verification, explicit threshold~\eqref{eq:family-line-only-counts}, and complete resource bounds \\
\bottomrule
\end{tabular}
\end{center}

\phantomsection\label{app:common-notation}
\paragraph{Common notation.}
\begin{center}
\small
\begin{tabular}{@{}P{.28\linewidth}@{\quad}P{.68\linewidth}@{}}
\toprule
Symbol & Meaning \\
\midrule
$\F,\F_Q$ & Code field of size $q$, and base field of the arithmetic tower \\
$D,g$ & Symbols per line and group order, with $D=Q+1$ in the arithmetic family \\
$N,K$ & Code length $D^2g$ and message dimension $k_Ak_Bg$ in the four-view construction \\
$M,v$ & Labelled lines $4Dg$ and views $4g$ \\
$W,y,S,S^*$ & Word fixed at registration in the soundness proof, partial observation $y\in\F^S$, observed positions $S$, and requested positions $S^*$ \\
$\ill(W),\ill_{\rm init}(y)$ & Failed-line count for $W$ and initialization charge computed from $y$ \\
$h,\kinit$ & Erasure threshold for a line and incompatible-line threshold at initialization \\
$\ill_{\max},R,u,t$ & Budgets for failed lines, heavy lines, missing coordinates, and observed disagreements \\
$d_{\min},d_C$ & Lower bounds on local and global minimum distance \\
$e,\Lambda_e$ & Degree cap and closure capacity in shared completion \\
$L_*,q_*,n_{\rm hole}$ & Per-side caps on heavy and non-heavy initial holes, and total cap $2(L_*+q_*)$ \\
$\mathsf C_e(s),\widehat{\mathsf T}$ & Closure cap for $s$ initial holes and bound on observed disagreements \\
$\ell,a_{\rm cl},n_{\rm dr},n_{\rm req}$ & Number of clients, draws per client, total draws, and sufficient draw threshold for the fixed family \\
$\lambda,\kappa,\Qhash$ & Security parameter, digest width in bits, and random-oracle query cap for the complete experiment \\
\bottomrule
\end{tabular}
\end{center}

%% file: sections/ai-disclosure.tex
\begin{samepage}
\section*{Disclosure of AI Assistance}
\phantomsection\label{sec:ai-disclosure}\label{app:ai}
The authors supplied the main ideas and approach and take full responsibility for the claims, proofs, and references.
GPT-5.6 Sol, GPT-6 Astra (both OpenAI), and Claude Opus 5.5 (Anthropic) substantially assisted with the development and organization of proofs and with manuscript drafting and revision.
Gemini (Google) assisted with English editing and manuscript organization.

\end{samepage}

%% file: sections/appendix-e-authentication/authentication-instance.tex
\section{Merkle Authentication}\label{app:authentication-instance}
All four labelled-line types share a coordinate-leaf Merkle tree. Fixation relates verified openings to a word fixed before sampling. Position binding bounds the probability that verified replies disagree on a shared coordinate.

\paragraph{Coordinate-leaf tree and openings.}
Leaves follow the public order on physical coordinates~\eqref{eq:public-leaf-order}.
As in~\cite[Appendix~B, Definition~15]{FRIDA}, each interval of $n>1$ leaves splits into subintervals of $\lceil n/2\rceil$ and $\lfloor n/2\rfloor$ leaves, giving height $d_{\rm tree}=\lceil\log_2N\rceil$ without padding.
The three Merkle input types use distinct domains and an injective encoding with unambiguous parsing.
A leaf input contains a coordinate label and canonical field value, and an internal-node input contains two ordered child digests.
The outer-root input contains the exact $\mathsf{ctx}$ bytes, $N$, the tree descriptor, and the inner root, and its digest is the registered commitment $\mathsf{com}$.

For the set $\mathsf U=S_i^*$ of coordinates requested by client~$i$, a node is occupied if its subtree meets $\mathsf U$. The canonical reply gives each requested leaf value once in increasing coordinate order and exactly one helper digest at each unoccupied child of an occupied internal node, ordered left to right. Draw labels determine the positions, so verification requires the exact ordered values and helpers, valid field encodings, and the recomputed outer root $\mathsf{com}$ under the expected registered context. Honest canonical replies verify because recomputing the committed tree reproduces $\mathsf{com}$.

\paragraph{Public context.}
The public context identifies the code field and family member, labelled ports, physical numbering, recovery and sampling parameters, and authentication format. In the fixed family it identifies the local checks chosen in Section~\ref{sec:local-preprocessing}. The clients and extractor verify against the same exact $\mathsf{ctx}$ bytes.

\subsection{Registration-time fixation and position binding}\label{app:merkle-guarantees}
Given the registration-time history $\mathsf{hist}$ (comprising all oracle queries and answers up to registration of $(\mathsf{ctx},\mathsf{com})$), we define the reference word $W=W(\mathsf{hist})\in\F^N$ as follows. For each coordinate in $[N]$, trace its path down from $\mathsf{com}$ using the first preimage recorded for each digest in the registration-time query log. If the log provides a complete path with the required input types, the exact context $\mathsf{ctx}$, and the expected coordinate labels, set that coordinate of $W$ to the leaf value; otherwise set it to zero.

A query is \emph{fresh} if its input is new.
Let $E_{\rm col}$ be the event that distinct oracle inputs collide.
The event $E_{\rm child}$ occurs when a fresh answer equals a child digest named in an earlier or current typed internal-node or outer-root input.
The event $E_{\rm root}$ occurs when the registered outer root has no preimage in the registration log and a fresh post-registration answer equals it.
Put $E_{\rm or}=E_{\rm col}\cup E_{\rm child}\cup E_{\rm root}$.
The cap $\Qhash$ counts every oracle call in the complete experiment, including adversarial queries, both verification passes, and failed or repeated attempts.

\Needspace{5\baselineskip}
\begin{lemma}[Merkle fixation and position binding]\label{lem:auth-instance}
For one registered pair and a $\kappa$-bit classical random oracle with at most $\Qhash$ complete-experiment queries,
\[
\begin{aligned}
 \Pr[E_{\rm or}]&\le\min\left\{1,\frac{3\Qhash^2+\Qhash}{2^\kappa}\right\},\\
 \Pr[E_{\rm col}]&\le\min\left\{1,\frac{\Qhash(\Qhash-1)}{2^{\kappa+1}}\right\}.
\end{aligned}
\]
Outside $E_{\rm or}$, every reply verified by a client or the extractor agrees with $W$, so $\Badval\subseteq E_{\rm or}$.
Outside $E_{\rm col}$, two verified canonical replies under the same registered pair agree on every shared coordinate, so $\mathsf{Bad}_{\rm bind}\subseteq E_{\rm col}$.
The adversary may be computationally unbounded subject to the classical oracle-query budget.
\end{lemma}
\begin{proof}
We adapt the query-log argument of~\cite[Appendix~B, Lemma~13]{FRIDA} to the typed tree and outer root. At the $j$th fresh query, the input and earlier answers are fixed when the oracle draws a uniform answer. It matches an earlier output with probability at most $(j-1)/2^\kappa$, giving the collision bound $\Qhash(\Qhash-1)/2^{\kappa+1}$. A fresh answer also has at most $2\Qhash$ child-digest targets named in earlier inputs or its own input, giving $\Pr[E_{\rm child}]\le2\Qhash^2/2^\kappa$. When the registered root has no preimage in the registration log, it supplies one fixed target for the fresh post-registration answers, so $\Pr[E_{\rm root}]\le\Qhash/2^\kappa$. Adding the three bounds gives the stated bound on $\Pr[E_{\rm or}]$.

For fixation, trace a verified opening down from the registered outer digest.
Resolving an outer digest missing from the registration log triggers $E_{\rm root}$, and replacing its recorded preimage triggers $E_{\rm col}$.
Outside these events, the recorded outer input fixes the inner root. Each fixed parent input then fixes both child digests.
Replacing a resolved child's preimage triggers $E_{\rm col}$, and a child's first resolution after registration triggers $E_{\rm child}$, since its digest was named in the parent input recorded at registration.
An ill-typed input recorded at registration would also have to be replaced for the opening to verify.
Thus outside $E_{\rm or}$, each verified leaf follows the same typed path as at registration and gives the corresponding coordinate of $W$.

For position binding, two verified replies under the same registered pair recompute the same outer digest. Outside $E_{\rm col}$, equal queried digests have identical inputs. Along the path to any shared requested leaf, equality of the parent input fixes the next child digest. At the leaf, the identical input fixes its coordinate label and value.
\end{proof}
The bounds in Lemma~\ref{lem:auth-instance} give $\eps_{\rm fix}$ and $\eps_{\rm bind}$, respectively, in Section~\ref{sec:auth-interface}.

\subsection{Multiproof sizes}\label{app:multiproof-sizes}
In the coordinate-leaf tree, a nonempty request $\mathsf U$ opens $s=|\mathsf U|$ leaves. Let $n_{\rm int}$ count occupied internal nodes and $n_{\rm help}$ count helper digests. The occupied proof subtree, completed with helpers, is full binary, so $s+n_{\rm help}=n_{\rm int}+1$. Hashing requested leaves, internal nodes, and the outer root gives
\begin{equation}\label{eq:multiproof-identity}
 n_{\rm help}(\mathsf U)=n_{\rm int}(\mathsf U)-s+1,
 \qquad V_{\rm hash}(\mathsf U)=2s+n_{\rm help}(\mathsf U).
\end{equation}
At depth $j$ there are at most $\min(2^j,s)$ occupied nodes. With $d_{\rm tree}=\lceil\log_2N\rceil$ and $j_s=\lfloor\log_2s\rfloor$,
\begin{equation}\label{eq:multiproof-frontier}
\begin{aligned}
 n_{\rm help}(\mathsf U)&\le\sum_{j=0}^{d_{\rm tree}-1}\min(2^j,s)-s+1\\
 &=2^{j_s+1}+(d_{\rm tree}-j_s-2)s.
\end{aligned}
\end{equation}
For $\mathsf U=\varnothing$, set $n_{\rm help}(\mathsf U)=V_{\rm hash}(\mathsf U)=0$. Draw labels determine the leaf addresses. A canonical reply to a nonempty request carries $s$ values and $\kappa n_{\rm help}$ helper bits.

Appendices~\ref{app:full-resource-ledger} and~\ref{app:cross-scheme-comparison} use the identity~\eqref{eq:multiproof-identity} and the frontier bound~\eqref{eq:multiproof-frontier} to account for verification hashes and reply sizes across both communication hops.

%% file: sections/appendix-a-code/ltc-appendix.tex
\section{Public Orders, Encoding, and Distance}\label{ltc:app-count}
\subsection{Public coordinate and message orders}\label{app:public-orders}
Fix a public bijection $\operatorname{idx}:G\to[g]$ and number physical coordinates by
\begin{equation}\label{eq:public-leaf-order}
 (x,i,j)\longmapsto D^2\operatorname{idx}(x)+Di+j.
\end{equation}
This order is used for serialization, requests, and Merkle leaves.
For the systematic ports of Section~\ref{sec:systematic-encoding}, the input symbol for $(x,I_A(\iota),I_B(\jmath))$ has message index
\begin{equation}\label{eq:public-message-order}
 (\iota k_B+\jmath)g+\operatorname{idx}(x).
\end{equation}

\subsection{Whole-kernel systematic extension}\label{app:systematic-extension}
\begin{proof}[Lemma~\ref{ltc:systematic}]
\phantomsection\label{proof:systematic-extension}
Let $\mathfrak a=\ker(\operatorname{aug}:\mathcal R\to\F)$. Jennings' theorem~\cite[Theorem~1.2]{Jennings} gives nilpotence of the augmentation ideal over the prime field. This property persists under scalar extension to $\F$. For a square matrix $P$ over $\mathcal R$ with invertible augmentation $P_0$, write $P_0^{-1}P=I+E$, where every entry of $E$ belongs to $\mathfrak a$. If $\mathfrak a^s=0$, then $E^s=0$, so
\[
P^{-1}=\left(\sum_{j=0}^{s-1}(-E)^j\right)P_0^{-1}.
\]
In particular, the selected pivot blocks $P_A$ of $\mathbf A$ and $P_B$ of $\mathbf B$ are invertible.

For any information rectangle, the two directional completions are therefore unique, using $P_A$ on the left and $P_B$ on the right.
After both solves, set $Y=\mathbf A\mathbf X$.
Solving the row constraints leaves the completed columns indexed by $I_B$ unchanged, so those columns of $Y$ vanish.
Writing $Y'$ for the submatrix of $Y$ on the remaining columns $[D]\setminus I_B$, the right equations give
\[
 Y'P_B=Y\mathbf B=\mathbf A(\mathbf X\mathbf B)=0.
\]
Invertibility of $P_B$ gives $Y'=0$ and hence $Y=0$, so the extension satisfies both defining systems. Any codeword with the same information rectangle is determined by the same two solves. Restriction to the information coordinates is therefore a bijection, and $\dim_{\F}\C=k_Ak_Bg$ since each of the $k_Ak_B$ free entries in $\mathcal R$ holds $g$ field symbols.
\end{proof}

\input{sections/appendix-a-code/central-quotient-solve}

\input{sections/appendix-a-code/distance-bounds}

%% file: sections/appendix-a-code/central-quotient-solve.tex
\subsection{Central-quotient solving}\label{app:central-quotient-solve}

\paragraph{Central order-$p$ solver.}
The solver uses a prepared chain of central order-$p$ quotients with kernel generators, transversals, fibre cycles, and labelled translations (constructed for the arithmetic family in Appendix~\ref{app:effective-central-tower}). We describe the left solve $P\phi=y$ for a sparse-monomial $c\times c$ matrix $P$ over $\mathcal R=\F[G]$ with invertible scalar augmentation. The same algorithm solves $P^{\mathsf T}\phi^{\mathsf T}=y^{\mathsf T}$ over the opposite ring $\mathcal R^{\rm op}$ for a right solve. Solver work is counted in code-field, array, and index operations.

\begin{proof}[Correctness of the recursive solver]
Induct on $n=|G|$. At $n=1$, solve using $\operatorname{aug}(P)^{-1}$. Otherwise choose the first central kernel $\langle z\rangle$ of order $p$, write $\xi=1-z$, and let $\pi:\mathcal R\to\F[G/\langle z\rangle]$ be the quotient map. Since $z$ is central of order $p$, we have $\xi\in Z(\mathcal R)$ and, in characteristic $p$, $\xi^p=1-z^p=0$. Moreover,
\[
 \ker\pi=\xi \mathcal R.
\]
On each fibre $(z^ix)_{0\le i<p}$, multiplication by $\xi$ sends coefficients $\psi_i$ to cyclic differences $\psi_i-\psi_{i-1}$.
For $\vartheta\in\ker\pi$, the coefficients $\vartheta_i=[\vartheta]_{z^ix}$ sum to zero on each fibre.
To solve $\xi\psi=\vartheta$, take $\psi_{p-1}=0$ and $\psi_i=\sum_{j=0}^i\vartheta_j$ for $0\le i<p-1$.
The cyclic differences equal $\vartheta_i$, including at $i=p-1$ by the zero sum.
This takes one prefix-sum pass over the fibres.

The quotient matrix $\pi(P)$ has the same invertible augmentation as $P$.
Starting with $s_0=y$, recursively solve $\pi(P)\bar\phi_j=\pi(s_j)$ and lift $\bar\phi_j$ to $\phi_j$ on the chosen transversal. At each $j=0,\ldots,p-1$, divide the residual, which has zero quotient image, to get the next right-hand side
\[
 \xi s_{j+1}=s_j-P\phi_j.
\]
Horner's rule assembles $\phi=\sum_{j=0}^{p-1}\xi^j\phi_j$ using cyclic differences to multiply by $\xi$. The residual equation and centrality of $\xi$ give
\[
 P\phi=\sum_{j=0}^{p-1}\xi^j(s_j-\xi s_{j+1})
       =s_0-\xi^ps_p=y.
\]
Lemma~\ref{ltc:systematic} gives invertibility of $P$, so the computed solution is unique.
In the opposite ring, $z$ remains central and the same fibres implement division by $\xi$.
The augmentation of $P^{\mathsf T}$ is invertible, so the argument also applies to the right solve.
\end{proof}

\paragraph{Encoding work and workspace.}
For a fixed $c\times c$ monomial system with $S_P$ nonzero monomials, the $p$ quotient solutions each occupy $cn/p$ packed coefficients, so their lifted monomial products cost $O(S_Pn)$ in total. Projection, residual formation, division on central fibres, and Horner assembly add $O(cpn)$ work. Thus the nonrecursive work is $O((S_P+cp)n)$ operations. Its recurrence and solution are
\[
 T_P(n)\le pT_P(n/p)+O((S_P+cp)n),\qquad
 T_P(n)=O(c^2n+(S_P+cp)n\log_p n).
\]
The encoder makes $k_B$ left solves and $D$ right solves. Each pivot block has at most $c_X^2$ monomials. With $c_{\max}=\max(2r_A,2r_B)$, the two scalar pivot inverses cost $O(c_{\max}^3)$ once, and the $O(c^2g)$ base costs sum to $O(Dc_{\max}^2g)$ over at most $2D$ solves. Since $c_{\max}\le D$ and $N=D^2g$, these costs contribute $O(Dc_{\max}^2+c_{\max}N)$.
Forming each right-hand side from the nonpivot block costs $O(c_Xk_Xg)$. With $m=\log_p g$, the total work is
\[
 O\!\left(Dc_{\max}^2+\left[c_{\max}+\frac{c_{\max}(c_{\max}+p)m}{D}\right]N\right)
\]
code-field and index operations.

At a level of size $n$, the packed solutions, residual, and accumulator occupy $O(c_{\max}n)$ cells. The active levels of a depth-first solve sum to $O(c_{\max}g)$ cells. The fixed monomial descriptions and the inverses of the augmentations use $O(c_{\max}D)$ more. Processing the right-hand sides sequentially gives $O(c_{\max}g+c_{\max}D)$ auxiliary field cells beyond the input, output, and prepared group data.

%% file: sections/appendix-a-code/distance-bounds.tex
\subsection{Distance bounds}\label{app:code-distance}

\begin{proof}[Distance bounds, Proposition~\ref{prop:distance}]
\phantomsection\label{proof:distance}
Let $W\ne0$ be a codeword with group-algebra matrix $\mathbf X$, so $\mathbf A\mathbf X=0$ and $\mathbf X\mathbf B=0$. The distance of $C_A^{\cap}$ makes every set of fewer than $d_A^{\cap}$ columns of $\operatorname{aug}(\mathbf A)$ independent.
On any such set of columns, choose a square submatrix of $\mathbf A$ with invertible augmentation. The submatrix is then invertible over $\F[G]$.
Applying the selected rows of $\mathbf A\mathbf X=0$ shows that every nonzero column of $\mathbf X$ is supported on at least $d_A^{\cap}$ row labels. By the symmetric argument with $\mathbf B$, each nonzero row has at least $d_B^{\cap}$ nonzero column entries. Thus there are at least $d_A^{\cap}$ disjoint nonzero row strips, each containing at least $d_B^{\cap}$ physical nonzero coordinates, proving $d(\C)\ge d_A^{\cap}d_B^{\cap}$.

For the relative-distance bound, fix $Z\in\{A,B\}$ and use the $Z$-port graph on two copies of $G$ with local constraints $C_Z^0,C_Z^1$. In a fixed-$i$ row strip, the labelled edge $j$ from $(0,x)$ to $(1,xb_j)$ carries $W(x,i,j)$. Its two endpoints impose the two $B$-typed row checks. In a fixed-$j$ column strip, the edge $i$ from $(0,x)$ to $(1,a_ix)$ carries the same coordinate and its endpoints impose the $A$-typed column checks.
Write $T_Z$ for the port averaging operator of Section~\ref{sec:distance-completion}. The normalized adjacency in part order $(0,1)$ is $M_Z=\left[\begin{smallmatrix}0&T_Z^*\\T_Z&0\end{smallmatrix}\right]$. The part-constant modes have eigenvalues $1$ and $-1$, and on their orthogonal complement the operator norm is at most $\sigma_Z$ by the bound on $T_Z$. Thus for $f\perp\mathbf 1$, the all-ones vector, the $-1$ mode contributes a nonpositive term and $\langle f,M_Zf\rangle\le\sigma_Z\|f\|^2$.
For a nonzero word on this graph, let $\Omega$ be the vertices incident to nonzero edges. Every vertex in $\Omega$ has at least $d_Z=\delta_ZD$ incident nonzero edges, and therefore at least that many neighbours in $\Omega$, counted with labels. Decompose $1_\Omega=(|\Omega|/(2g))\mathbf 1+f$. The degree bound and spectral estimate give
\[
 \delta_Z|\Omega|\le\langle1_\Omega,M_Z1_\Omega\rangle
 \le\sigma_Z|\Omega|+(1-\sigma_Z)|\Omega|^2/(2g).
\]
As $\delta_Z>\sigma_Z$, this yields $|\Omega|\ge2g(\delta_Z-\sigma_Z)/(1-\sigma_Z)$. Counting nonzero incident edges at both endpoints gives weight at least $d_Z|\Omega|/2\ge\gamma_ZDg$ with $\gamma_Z=\delta_Z(\delta_Z-\sigma_Z)/(1-\sigma_Z)$. Each nonzero row strip therefore has weight at least $\gamma_B Dg$, and the disjoint-strip count gives $d(\C)\ge d_A^{\cap}\gamma_BDg$. Interchanging $A$ and $B$ yields the other bound, and division by $N=D^2g$ proves~\eqref{ltc:delta}.
\end{proof}

%% file: sections/appendix-a-code/completion-proofs.tex
\section{Shared Completion, Local Testing, and Partial Recovery}\label{app:completion-testing}
We prove the shared completion guarantee, then use it to analyze full-word testing and recovery from partial observations.
\subsection{Tensor lifting}\label{app:tensor-lifting}
\begin{proof}[Lemma~\ref{lem:tensor-lift}]
Deletion of fewer than the respective code distances makes $C_A\to C_A|_{\bar I}$ and $C_B\to C_B|_{\bar J}$ injective, hence isomorphisms onto their images. Extending bases of those images to ambient bases identifies the row--column intersection in Lemma~\ref{lem:tensor-lift} with $(C_A|_{\bar I})\otimes(C_B|_{\bar J})$. Restriction from $C_A\otimes C_B$ is the tensor product of the two puncturing isomorphisms, and the tensor product of their inverses supplies its unique lift. The same injectivity identifies each supplied legal row and column with its counterpart in the lift.
\end{proof}

\subsection{Closure bounds and filling}\label{app:hole-closure}

\paragraph{View-graph edge bound.}\label{app:view-edge-bound}
Let $P_H$ be the normalized adjacency operator of the $2D$-regular view multigraph, and write $v=4g$ and $\sigma=\sigma_{\rm avg}$.
On type-constant functions, $P_H$ averages the two bit flips and has eigenvalues $1,0,0,-1$.
On functions of zero mean on each type, its Rayleigh quotient is at most $\sigma$ by the directional norm bounds.
Since each labelled translation permutes $G$, the two subspaces are invariant and orthogonal. The nonconstant type-constant modes have eigenvalues at most zero, so the Rayleigh quotient on the orthogonal complement of the global constant is at most $\sigma$.
The internal edge count is $|E(\mathcal U)|=D\langle1_{\mathcal U},P_H1_{\mathcal U}\rangle$.
For $n=|\mathcal U|$, the orthogonal part of $1_{\mathcal U}$ has squared norm $\|1_{\mathcal U}-(n/v)\mathbf 1\|^2=n-n^2/v$, giving~\eqref{ltc:mix}.

\begin{proof}[Closure growth, Lemma~\ref{lem:hole-completion}]
\phantomsection\label{proof:hole-closure}
For an empty seed the closure is empty. Otherwise put $a=|\mathcal Z_0|$. Each added vertex has at least $e+1$ edges to the current holes, so at every reached size $n$,
\[
 (e+1)(n-a)\le |E(\mathcal Z_n)|\le\mu_{\rm mix}(n).
\]
Since $2a<\Lambda_e$, $\mu_{\rm mix}(2a)<(e+1)a$. The process adds one vertex at a time, so it stops before reaching $2a$. At termination every assigned view has at most $e$ edges into the holes by the closure rule.
\end{proof}

\begin{proof}[Filling, Lemma~\ref{lem:hole-completion}]
\phantomsection\label{proof:hole-filling}
After closure, each assigned view has at most $e$ labelled edges into the unassigned set $\mathcal U=\mathcal Z$. Maintain two invariants while filling: assigned matrices are legal and agree along all shared lines, and every assigned view has at most $e$ labelled edges into $\mathcal U$. Both hold initially. If $\mathcal U\ne\varnothing$, then $|\mathcal U|\le|\mathcal Z|<2|\mathcal Z_0|<\Lambda_e$, and~\eqref{ltc:mix} gives
\[
 \frac{2|E(\mathcal U)|}{|\mathcal U|}
 \le2D\sigma+2D(1-\sigma)|\mathcal U|/v<e+1.
\]
There is therefore a view $z$ incident to at most $e$ edges within $\mathcal U$. Let $I$ and $J$ be its missing row and column labels, with $|I|+|J|\le e<d_{\min}$. Assigned neighbours supply legal full rows indexed by $\bar I$ and legal full columns indexed by $\bar J$. Write $X_{\rm row},X_{\rm col}$ for their restrictions to $\bar I\times\bar J$. Their difference belongs to
\[
 (C_A^{s_A}|_{\bar I})\otimes\F^{\bar J}
  +\F^{\bar I}\otimes(C_B^{s_B}|_{\bar J}).
\]
For a retained position $(i,j)$, the two suppliers and $z$ are three corners of the same physical coordinate. If its fourth, diagonally opposite view is assigned, shared-line agreement makes the suppliers' values equal. Otherwise that opposite view lies in $\mathcal U$. For a fixed retained row, each disagreement uses a distinct labelled column edge from its supplier to a hole. The supplier already uses one of its at most $e$ edges into the holes for the row it shares with $z$. This leaves at most $e-1$ disagreements. By symmetry each retained column has at most $e-1$. Labels distinguish the edges even when the translated group elements coincide. Property $J_e$ forces $X_{\rm row}=X_{\rm col}$.

Lemma~\ref{lem:tensor-lift} now gives the unique legal tensor view extending this rectangle and preserving every supplied row and column. Assign it to $z$ and delete $z$ from $\mathcal U$. The new view agrees with all assigned neighbours and has at most $e$ edges into the smaller hole set. The boundary degrees of earlier assignments can only decrease. Thus the invariants continue until $\mathcal U$ is empty. The four occurrences of each physical coordinate then agree around its square of shared lines, defining a word $c\in\C$. Each view outside $\mathcal Z$ retains its initial assignment, proving the preservation assertion of Lemma~\ref{lem:hole-completion}.
\end{proof}

\subsection{Local testing on full words}\label{app:testing-proof}
\begin{proof}[Local testability, Theorem~\ref{thm:testing}]
\phantomsection\label{proof:local-testability}
For the uniform defining-line test put $\beta(W)=\ill(W)/M$. Set
\begin{equation}\label{ltc:constants}
 L_{\rm r}=\frac{d_{\min}-1}{D}+\frac{4D}{d_{\min}},\qquad
 \beta_0=\min\left\{\frac{d_{\min}\Lambda_e}{8M},\frac{1-K/N}{L_{\rm r}}\right\}.
\end{equation}
These constants are positive and depend only on the local and spectral data, since $K/N=k_Ak_B/D^2$ and $\Lambda_e/M=((e+1)/(2D)-\sigma_{\rm avg})/(D(1-\sigma_{\rm avg}))$.

For a complete word $W$, let $I_z,J_z$ be the illegal row and column labels in its view $z$, with counts $i_z,j_z$. If either count is at least $d_{\min}$, mark $z$ as a forced hole. Otherwise the retained rows and columns are legal and agree on their intersections (both equal the entries of $W_z$). Tensor lifting supplies a unique legal $Y_z$ preserving them, with $E_z=W_z-Y_z$ supported on $I_z\times J_z$. Let $F$ be the forced holes and $V_{\rm corr}$ the other, corrected, views.

On $V_{\rm corr}$ form the bipartite conflict multigraph of shared labelled lines where corrected endpoint matrices disagree. For a conflicting row line, the difference of the two legal rows is a nonzero word of the same local $B$-code and is supported on the union of the endpoints' illegal-column sets. Hence the endpoints have at least $d_{\min}$ illegal-column incidences. A conflicting column line gives the same bound on illegal-row incidences. Let $U$ be a minimum vertex cover and set $\mathcal Z_0=F\mathbin{\dot\cup}U$. Outside $\mathcal Z_0$, all assigned views are legal and adjacent assignments agree.

By K\"onig's theorem the size of $U$ equals a maximum matching size. The matched edges have disjoint endpoints outside $F$, and each charges at least $d_{\min}$ illegal-line incidences. Every forced view also charges at least $d_{\min}$. Each illegal labelled line has two view endpoints, so
\[
 d_{\min}|\mathcal Z_0|\le\sum_z(i_z+j_z)=2\ill(W).
\]
For a corrected view, $i_z,j_z\le d_{\min}-1$ and $2i_zj_z\le(d_{\min}-1)(i_z+j_z)$. Summing gives
\begin{equation}\label{ltc:coarse}
 |\mathcal Z_0|\le\frac{2\ill(W)}{d_{\min}},\qquad
 \sum_{z\notin\mathcal Z_0}\wt(E_z)\le(d_{\min}-1)\ill(W).
\end{equation}
If $\beta(W)\le\beta_0$, then $|\mathcal Z_0|\le2M\beta_0/d_{\min}\le\Lambda_e/4$. Shared completion applies, with $|\mathcal Z|<2|\mathcal Z_0|$ for a nonempty seed and $\mathcal Z=\varnothing$ otherwise. It produces $c\in\C$ agreeing with $Y_z$ outside $\mathcal Z$. Each changed physical coordinate has four view corners. Outside $\mathcal Z$ the matrices $E_z$ count these corners. Inside $\mathcal Z$ there are at most $D^2|\mathcal Z|$ corners in total. Thus
\[
 \begin{aligned}
 4\wt(W-c)&\le(d_{\min}-1)\ill(W)+D^2|\mathcal Z|\\
 &\le\left(d_{\min}-1+\frac{4D^2}{d_{\min}}\right)\ill(W).
 \end{aligned}
\]
Using $M=4N/D$ proves $\dist(W,c)\le L_{\rm r}\beta(W)$. For $\beta(W)\le\beta_0$, this implies $\dist(W,\C)\le L_{\rm r}\beta(W)\le(1-K/N)\beta(W)/\beta_0$. For $\beta(W)>\beta_0$, systematic completion of the $K$ information symbols gives a codeword differing in at most $N-K$ coordinates, hence the same bound on $\dist(W,\C)$ holds. Rearranging yields the claimed tester bound, with perfect completeness and exactly $D$ symbol queries.
\end{proof}

%% file: sections/appendix-b-recovery/recovery-supplement.tex
\subsection{Initialization for partial recovery}\label{app:recovery-initialization-proof}\label{app:recovery}
For recovery from partial observations, initialization supplies the legal, agreeing views required for shared completion.
\begin{proof}[Initialization comparison, Lemma~\ref{lem:recovery-initialization}]
\phantomsection\label{proof:initialization-comparison}
An assigned view is non-heavy. Its checked tensor lift is legal and preserves every retained completed row and column. A discrepancy with an observed position can therefore occur only where an omitted row meets an omitted column. If two assigned views share a compatible line, they preserve its unique legal completion. On an incompatible shared line, the legal rows or columns that the two assigned views place on this line can differ only at its at most $h$ missing positions and at the other incompatible crossing labels at the two endpoints. The shared line accounts for one of at most $\kinit$ incompatible labels at each endpoint. Their difference thus has weight at most $h+2(\kinit-1)<d_{\min}$ and is a word of the corresponding component code, so it vanishes. Assigned neighbours agree on all shared lines.

Now fix $S\subseteq[N]$, $y\in\F^S$, and an arbitrary complete extension $W\in\F^N$ of $y$. Every incompatible non-heavy line is illegal in $W$. Consider a non-heavy view with at most $\kinit$ illegal incident lines of $W$.
To construct a witness lift, omit all lines illegal in $W$. This removes at most $\kinit<d_{\min}$ labels, and the remaining rows and columns agree at their intersections because they come from $W$.
Lemma~\ref{lem:tensor-lift} supplies a legal view $Y^*$ preserving these lines. Uniqueness of completion gives the row or column of $W$ on every legal incident line. On an illegal but compatible retained row, the completed row and the row of $Y^*$ can disagree only at erased positions or at crossings with other illegal lines. There are at most $h$ of the first kind and $\kinit-1$ of the second. Both rows lie in the same component code, so the distance bound makes them equal. The same argument applies to columns. Every retained completed line therefore agrees with $Y^*$ and the checked lift succeeds.

Thus, for every complete extension $W$ of $y$, each non-heavy initial hole, including one whose checked lift failed, is incident to at least $\kinit+1$ lines illegal in $W$.
At an assigned view $z$, all $\inc_z$ incompatible incident lines are illegal in $W$. Every illegal labelled line has exactly one incidence on each bipartite side. Thus each side charge in~\eqref{eq:initialization-budget}, and $\inc(y)$ itself, is at most $\ill(W)$. Taking the maximum proves $\ill_{\rm init}(W|_S)\le\ill(W)$ for every complete $W$ and every $S$.

Finally, suppose $\kinit\le e$ and $h\le e-1$. At a non-heavy view with at most $\kinit$ incompatible labels, the retained completed rows and columns give two rectangles whose difference belongs to the punctured tensor-sum space in Definition~\ref{ltc:je}. Both rectangles agree with the observations where present, so every row and column of their difference has at most $h\le e-1$ nonzeros. Property $J_e$ forces equality of the rectangles and the checked lift succeeds. Therefore a non-heavy initial hole has more than $\kinit$ incompatible incident lines. Each side charge in~\eqref{eq:initialization-budget} is at most $\inc(y)$. The maximum includes $\inc(y)$ itself, so $\ill_{\rm init}(y)=\inc(y)$.
\end{proof}

\input{sections/appendix-b-recovery/recovery-budgets}

\subsection{Verification of the recovery interface}\label{app:instance-interface-proof}
\begin{proof}[Corollary~\ref{cor:instance-interface}]
The construction of Section~\ref{sec:source-hypotheses} and Theorem~\ref{thm:encoding} give a systematic code defined exactly by its line checks, and $d_C$ is a distance lower bound by hypothesis.
The configuration of Section~\ref{sec:recovery-algorithm} gives nonnegative $(h,\ill_{\max},R,u,t)$ with $\ill_{\max}<M$ and $u<N$.
The recovery guarantee and checked outputs are the two guarantees of Theorem~\ref{thm:recovery}.
The upper bound on $t$ in~\eqref{eq:admissible-basic} gives $2(t+u)<d_C$.
\end{proof}

%% file: sections/appendix-b-recovery/recovery-budgets.tex
\subsection{Closure caps and observed disagreements}\label{app:recovery-budgets}

\paragraph{Integer closure bound.}\label{app:integer-closure-bound}
Consider the cap $\mathsf C_e(s)$ defined in Section~\ref{sec:recovery-parameters}. For $s=0$ the closure is empty. For $s>0$, the integer $s$ itself is feasible. If a seed of size $a\le s$ has closure size $n\ge s$, then $n<2a\le2s$ and Lemma~\ref{lem:hole-completion} gives $(e+1)(n-s)\le(e+1)(n-a)\le\mu_{\rm mix}(n)$. Thus $n\le\mathsf C_e(s)$ in this case, and the same bound holds for $n<s$ because $\mathsf C_e(s)\ge s$.

For consecutive integer caps $s,s+1$ in the domain, $\mathsf C_e(s+1)\ge\mathsf C_e(s)+1$. At $s=0$, this follows from $\mathsf C_e(0)=0$ and $\mathsf C_e(1)=1$. For $s>0$, if $n$ is feasible for $s$, then $n+1$ lies in $[s+1,2(s+1))$. The left side stays the same, and $\mu_{\rm mix}$ is nondecreasing because $0\le\sigma_{\rm avg}<1$.

\begin{proof}[Observed disagreement bound, Lemma~\ref{lem:residual-budget}]
\phantomsection\label{proof:observed-disagreements}
Fix $y\in\F^S$ satisfying the hypotheses of Lemma~\ref{lem:residual-budget}.
The initialization comparison gives legal, agreeing assigned views.
The counts in Section~\ref{sec:recovery-parameters} give $H_0\le2L_*$ and $I_0\le2q_*$. The hypothesis $2n_{\rm hole}=2(2L_*+2q_*)<\Lambda_e$ puts every sum in this budget rectangle in the domain of $\mathsf C_e$.
Lemma~\ref{lem:hole-completion} therefore applies to Steps~3--4 of Figure~\ref{fig:recovery-algorithm}, yielding the codeword $c$.
The closure contains at most $\mathsf C_e(H_0+I_0)$ views.
Each of its $H_0$ heavy initial holes has at least $h+1$ erased corners, giving at most $D^2\mathsf C_e(H_0+I_0)-(h+1)H_0$ observed corners in the closure.

A view $z$ outside the closure retains its initial assignment. Its disagreements can occur only where its incompatible rows and columns cross (Lemma~\ref{lem:recovery-initialization}). An incompatible row label and column label share at most one view because their edges flip different type bits, so their pair gives at most one corner. Thus there are at most $\inc(y)^2/4\le\ill_{\max}^2/4$ disagreement corners outside the closure. We can also count them by view. With $\inc_z\le\kinit$ incompatible lines at $z$, there are at most $\inc_z^2/4\le\kinit\inc_z/4$ crossings. The two side charges in~\eqref{eq:initialization-budget} give $\sum_{z\notin\mathcal Z}\inc_z\le2\ill_{\max}-(\kinit+1)I_0$. This gives the second bound in the minimum of~\eqref{eq:recovery-corner-bound}.

Each observed disagreement contributes four corners, so $4\wt(c|_S-y)\le F(H_0,I_0)$. In the budget rectangle, increasing either count by one increases the closure term by at least $D^2$ by the \hyperref[app:integer-closure-bound]{integer closure bound above}. Increasing $H_0$ increases the erased-corner deduction by $h+1$, and increasing $I_0$ decreases the outside bound by at most $\kinit(\kinit+1)/4$. Both losses are at most $D^2$ because local consistency gives $h+1,\kinit\le d_{\min}\le D$. Thus $F(H_0,I_0)\le F(2L_*,2q_*)$. Division by four and rounding down give~\eqref{eq:recovery-residual}.
\end{proof}

%% file: sections/appendix-a-code/family-and-testing.tex
\section{Construction of a Qualified Fixed-Field Family}\label{app:local-arithmetic}\label{app:arithmetic-family}
We first give a finite qualification certificate and choose fixed local proportions for any target rate, proving Lemma~\ref{lem:qualified-local-checks}. We then construct the arithmetic tower and assemble the uniform code family.
\input{sections/appendix-a-code/sparse-count}
\input{sections/appendix-a-code/fullrank-asymptotic}

%% file: sections/appendix-a-code/sparse-count.tex
\subsection{Finite local qualification}\label{app:fullrank-asymptotic}\label{app:sparse-count}
For a prime power $q\ge2$, let $B_q(l,c)=\sum_{j=0}^{\min(l,c)}\binom lj(q-1)^j$ when $c\ge0$, and $B_q(l,c)=0$ otherwise.
\begin{lemma}[Sparse matrix count]\label{ltc:joint}
For $1\le n_1,n_2\le D$, $w\ge1$, and $1\le v\le\min(n_1,n_2)$, at most
\[
 D^v B_q(D,w)^{\min\{2v,n_1,n_2\}}
\]
rank-$v$ matrices in $\F_q^{n_1\times n_2}$ have at most $w$ nonzero entries in each row and column.
\end{lemma}
\begin{proof}
Transpose to make $n_1\le n_2$ and put $B=B_q(D,w)$. Choose $v$ independent sparse rows in at most $D^vB^v$ ways. Choose pivot columns deterministically from these rows. The $v$ pivot columns, with at most $B^v$ choices in total, then determine the coefficients of every remaining row. Alternatively record all $n_1$ rows in at most $B^{n_1}$ ways. Taking the smaller bound proves the claim.
\end{proof}

%% file: sections/appendix-a-code/fullrank-asymptotic.tex
\paragraph{Qualification certificate.}
We use the sparse-matrix count to bound failures of $J_e$ for random local checks. The following certificate also includes the distance requirements.
For integers $2r<D$, $2\le e\le r$, $e<d\le D$, and $1\le d_{\cap}\le D$, put
\[
 w=e-1,\qquad T_D=\max\left\{1,\left\lfloor\frac{2d-2e-1}{e-1}\right\rfloor\right\},
 \qquad D_{\rm ret}=D-\lceil e/2\rceil.
\]
For real $\mathsf r\ge\mathsf e\ge0$ and $x\ge0$, define
\begin{equation}\label{eq:fullrank-continuous}
 F_{\mathsf r,\mathsf e}(x)=(\mathsf r-\mathsf e)x-
 \frac{(x-\mathsf e)_+^2-(x-(2\mathsf r-\mathsf e))_+^2}{4}.
\end{equation}
This function is nondecreasing and concave on $[0,\infty)$. Define
\[
\begin{aligned}
 \mathcal E_D(x)&=F_{r,e}(x)-x\log_q(2D)
 -\min\{2x,D_{\rm ret}\}\log_q B_q(D,e-1),\\
 \Xi_D&=\min_{x\in\{T_D,D_{\rm ret}/2,D_{\rm ret}\}\cap[T_D,D_{\rm ret}]}
 \mathcal E_D(x),
\end{aligned}
\]
with $\min\varnothing=+\infty$. Put
\begin{equation}\label{eq:fullrank-finite-j}
 P_J=\frac{52D_{\rm ret}}{q-1}\binom{2D}e q^{-\Xi_D},
 \qquad P_J=0\quad(T_D>D_{\rm ret}).
\end{equation}
\begin{equation}\label{eq:fullrank-finite-setup}
 \eps_D=\min\left\{1,\frac{4[B_q(D,d-1)-1]}{(q-1)q^r}
 +\frac{2[B_q(D,d_{\cap}-1)-B_q(D,d-1)]_+}{(q-1)q^{2r}}+P_J\right\}.
\end{equation}

\begin{lemma}[Qualification of full-rank local checks]\label{lem:finite-local-qualification}
Let $2r<D$, $2\le e\le r$, $e<d\le D$, and $1\le d_{\cap}\le D$. Draw two independent stacks uniformly from the full-row-rank $2r\times D$ matrices and split each into its first and last $r$ rows. The probability that any typed distance is below $d$, either intersection distance is below $d_{\cap}$, or any typed pair fails $J_e$ is at most $\eps_D$ in~\eqref{eq:fullrank-finite-setup}.
\end{lemma}
\begin{proof}
For a sparse witness to the failure of $J_e$, the typed distances restrict its possible rank.
A fixed witness \emph{survives} if it belongs to the tensor sum of Definition~\ref{ltc:je} for the sampled checks.
We bound survival probabilities in the independent-entry model, then transfer the union bound by conditioning on full row rank.
The same conditioned distribution gives the distance-failure terms.

For a matrix $H$, write $H|_T$ for its submatrix of columns indexed by $T$.
For a typed pair $(C_A^{s_A},C_B^{s_B})$ with deletion sets $I,J$, choose $N_I$ from the deleted block $H_A^{s_A}|_I$ alone, with rows spanning that block's left kernel.
A vector on $\bar I$ extends to a codeword exactly when its syndrome under $H_A^{s_A}|_{\bar I}$ lies in the column space of the deleted block.
Thus $\widehat H_A=N_I(H_A^{s_A}|_{\bar I})$ has kernel $C_A^{s_A}|_{\bar I}$, and the analogous $\widehat H_B$ has kernel $C_B^{s_B}|_{\bar J}$.
The tensor sum in Definition~\ref{ltc:je} is the kernel of the linear map $Z\mapsto\widehat H_A Z\widehat H_B^{\mathsf T}$.

The event $\mathsf{Good}_{\rm all}$, on which all typed distances are at least $d$, restricts which witness ranks we must count.
Given a nonzero witness to the failure of $J_e$, enlarge the deletion sets $I,J$ until $n_I+n_J=e$, where $n_I=|I|$ and $n_J=|J|$.
Choose the additional deletions to preserve one nonzero entry.
Write the resulting witness as $Z=LR^{\mathsf T}$ with rank $v$ and both factors of full column rank.
If $T_D=1$, the lower rank cutoff is automatic.
Otherwise $d>2e-1$, and the witness equation reads
\[
 (\widehat H_A L)(\widehat H_B R)^{\mathsf T}=0.
\]
Write $v_A=\rank(\widehat H_A L)$ and $v_B=\rank(\widehat H_B R)$. The two row spaces are orthogonal in $\F_q^v$, so $v_A+v_B\le v$. If $v_A=0$, every column of $Z$ lies in $\ker\widehat H_A$ with weight at most $w<d-n_I$, forcing $Z=0$ by the punctured distance bound. Thus $v_A,v_B\ge1$ by symmetry. The syndromes of $v_A+1$ independent witness columns are dependent, giving a nonzero punctured codeword of weight at most $(v_A+1)w$. Puncturing reduces distance by at most $n_I$. Applying the same argument to rows gives
\[
 d-n_I\le(v_A+1)w,\qquad d-n_J\le(v_B+1)w,
 \qquad 2d-e\le(v+2)(e-1).
\]
When $T_D>1$, this gives $v\ge\lceil(2d-e)/(e-1)\rceil-2=\lfloor(2d-2e)/(e-1)\rfloor\ge T_D$. In all cases the matrix dimensions give $v\le D-\max(n_I,n_J)\le D_{\rm ret}$.

To bound survival of a fixed witness of an allowed rank, use the independent-entry model and condition only on the deleted columns. Choose $a=r-n_I$ and $b=r-n_J$ independent vectors from the respective left kernels of these blocks. The retained columns remain uniform and independent. Taking these row combinations and multiplying by the full-column-rank factors $L,R$ gives independent uniform $U\in\F_q^{a\times v}$ and $V\in\F_q^{b\times v}$. Witness survival requires $UV^{\mathsf T}=0$. Choosing $i$ independent columns and expressing the rest in their span gives at most $\binom vi q^{ai+i(v-i)}$ rank-$i$ matrices $U$, each with orthogonality probability $q^{-bi}$. Thus
\begin{equation}\label{eq:fullrank-probability}
 \Pr[UV^{\mathsf T}=0]\le\sum_{i=0}^{\min(a,v)}\binom vi
 q^{-[(a-i)(v-i)+bi]}\le2^v q^{-F_{r,e}(v)},
\end{equation}
where $F_{r,e}$ is defined in~\eqref{eq:fullrank-continuous}.
Minimizing the exponent over real $i\in[0,\min(a,v)]$ lower-bounds the minimum over integer ranks. The real minimum is $\min(a,b)v$, $(a+b)v/2-(v^2+(a-b)^2)/4$, and $ab$ as $v$ lies below $|a-b|$, between $|a-b|$ and $a+b$, or above $a+b$. At fixed $a+b=2r-e$, these minima are smallest at $|a-b|=e$, giving~\eqref{eq:fullrank-continuous}.

For each typed pair and padded deletion choice, Lemma~\ref{ltc:joint} and~\eqref{eq:fullrank-probability} bound the probability that $\mathsf{Good}_{\rm all}$ holds and a rank-$v$ witness survives by $q^{-\mathcal E_D(v)}$ in the independent-entry model. Concavity on either side of $D_{\rm ret}/2$ gives $\mathcal E_D(v)\ge\Xi_D$ for every allowed rank. There are four typed pairs and $\binom{2D}e$ padded deletion choices. Count each nonzero witness up to multiplication by a nonzero scalar, dividing the count by $q-1$.

Since $q\ge2$, an independent-entry $2r\times D$ stack over $\F_q$ has full row rank with probability at least $\prod_{j\ge1}(1-q^{-j})\ge\prod_{j\ge1}(1-2^{-j})\ge(1/2)(3/4)(1-\sum_{j\ge3}2^{-j})=9/32$.
By independence, both stacks have full row rank with probability at least $(9/32)^2>1/13$.
Conditioning on this event therefore multiplies the upper bound on the joint event by at most $13$, giving the bound $P_J$ in~\eqref{eq:fullrank-finite-j}.
In the conditioned distribution, invariance under the right action of $\mathrm{GL}_D(\F_q)$ makes the typed and intersection kernels uniform among subspaces of dimensions $D-r$ and $D-2r$, respectively. A fixed nonzero vector lies in a uniform codimension-$r'$ kernel with probability $(q^{D-r'}-1)/(q^D-1)<q^{-r'}$. Count scalar orbits of low-weight words, and count intersection failures only when all typed distances are at least $d$. This gives $\eps_D$ in~\eqref{eq:fullrank-finite-setup}.
\end{proof}
If $\eps_D<1$, exhaustive qualification or rejection sampling followed by the stated tests selects fixed checks. Rejection sampling takes at most $(1-\eps_D)^{-1}$ attempts in expectation.

\subsection{Strict fixed-field margins}\label{app:fixed-field-checks}
We next give fixed-proportion conditions under which the finite qualification bound decays exponentially with $D$.
Write $H_2(t)=-t\log_2t-(1-t)\log_2(1-t)$ and $h_q(t)=[H_2(t)+t\log_2(q-1)]/\log_2q$, with endpoint continuity. For $0<\bar e<\bar r<1/2$ and $\bar d>2\bar e$, put
\[
\begin{aligned}
 H_{\bar e}&=h_q(\bar e),&m_0&=\bar r-\bar e-2H_{\bar e},\\
 \ell_{\rm ret}&=1-\bar e/2,&T_\infty&=\lfloor2\bar d/\bar e-2\rfloor,\\
 L_{\rm del}&=2H_2(\bar e/2)/\log_2q.
\end{aligned}
\]
\begin{lemma}[Fixed-field local qualification]\label{lem:sparse-local-existence}
Fix $0<\bar e<\bar r<1/2$, $2\bar e<\bar d<1-1/q$, and $0<\bar d_{\cap}<1-1/q$, with
\[
 h_q(\bar d)<\bar r,\quad h_q(\bar d_{\cap})<2\bar r,\quad
 F_{\bar r,\bar e}(\ell_{\rm ret}/2)>\ell_{\rm ret}H_{\bar e},\quad
 T_\infty m_0>L_{\rm del}.
\]
For $r=\lfloor\bar rD\rfloor$, $e=\lfloor\bar eD\rfloor$, $d=\lfloor\bar dD\rfloor$, $d_{\cap}=\lfloor\bar d_{\cap}D\rfloor$, and any
\[
 0<c_{\rm setup}<\min\{\bar r-h_q(\bar d),\ 2\bar r-h_q(\bar d_{\cap}),\ T_\infty m_0-L_{\rm del}\},
\]
we have $\eps_D\le q^{-c_{\rm setup}D}$ for all sufficiently large $D$.
\end{lemma}
\begin{proof}
The rounded parameters eventually satisfy the finite hypotheses, and $D_{\rm ret}=\ell_{\rm ret}D+O(1)$.
For all sufficiently large $D$, we also have $T_D=T_\infty$.
If $2\bar d/\bar e-2$ is nonintegral, convergence of the defining ratio gives the equality.
If $2\bar d/\bar e-2=c\in\mathbb Z$, then $c\ge3$ and $2d-(c+2)e+c-1\ge c-3\ge0$, while the defining cutoff ratio is eventually below $c+1$.
Hamming-ball entropy gives $B_q(D,e-1)\le q^{DH_{\bar e}}$.
Since $T_\infty$ is fixed and eventually $T_\infty\le e$, we obtain
\[
 \mathcal E_D(T_D)\ge T_\infty m_0D-O(\log D).
\]
Homogeneity of $F$ and rounding, at $\theta\in\{1/2,1\}$, give
\[
 \mathcal E_D(\theta D_{\rm ret})\ge
 D^2\bigl[F_{\bar r,\bar e}(\theta\ell_{\rm ret})-\ell_{\rm ret}H_{\bar e}\bigr]-O(D\log D).
\]
The bracket is strictly positive at $\theta=1/2$ by hypothesis and at $\theta=1$ by monotonicity of $F$. Thus the minimum $\Xi_D$ is at least $T_\infty m_0D-O(\log D)$. Finally $\binom{2D}e\le q^{L_{\rm del}D}$ makes $P_J\le q^{-(T_\infty m_0-L_{\rm del})D+O(\log D)}$. The two distance terms in~\eqref{eq:fullrank-finite-setup} are at most $q^{-D(\bar r-h_q(\bar d))+O(1)}$ and $q^{-D(2\bar r-h_q(\bar d_{\cap}))+O(1)}$, respectively. The strict choice of $c_{\rm setup}$ absorbs rounding and polynomial factors.
\end{proof}

\begin{proof}[Lemma~\ref{lem:qualified-local-checks}, target-rate choice]
\phantomsection\label{app:target-rate-choice}
For any fixed $\F_q$ and $\rho\in(0,1)$, choose $\bar r=(1-\sqrt\rho)/2$ and $\bar d=\bar d_{\cap}=4\bar e$. As $\bar e\downarrow0$, $T_\infty=6$, $H_{\bar e},L_{\rm del}\to0$, $m_0\to\bar r$, and $F_{\bar r,\bar e}(\ell_{\rm ret}/2)\to\bar r^2-(\bar r-1/4)_+^2>0$. Hence all strict inequalities of Lemma~\ref{lem:sparse-local-existence} hold for sufficiently small $\bar e>0$. For all sufficiently large $D$, the rounded parameters satisfy $2r<D$, $2\le e\le r$, and $e<d\le D$, and that lemma gives $\eps_D\le q^{-cD}$ for some $c>0$. Lemma~\ref{lem:finite-local-qualification} yields the claimed checks, and rounding gives $(1-2r/D)^2\ge(1-2\bar r)^2=\rho$.
\end{proof}

%% file: sections/appendix-b-carrier/arithmetic-carrier.tex
\subsection{Common arithmetic setup}\label{app:arithmetic-setup}
The same arithmetic setup supplies the port norm bound and the effective central quotient chain.

Fix $Q=p^a$ with an integer $a\ge1$, and put $D=Q+1$. We charge operations over the arithmetic field $\F_Q$ and the code field $\F_q$ separately. Choose the basis $1,\omega$ of $\F_{Q^2}$ over $\F_Q$. For odd $p$, take $\omega^2=c_{\rm ar}$ with $c_{\rm ar}$ nonsquare. For $p=2$, take $\omega^2+\omega+c_{\rm ar}=0$ with $\operatorname{Tr}_{\F_Q/\F_2}(c_{\rm ar})=1$. Let $\bar x=x^Q$ on $\F_{Q^2}$ and extend this automorphism by fixing $T_{\rm ar}$. Put $\mathcal K=\F_Q(T_{\rm ar})$ and
\[
 \mathcal D=\F_{Q^2}(T_{\rm ar})\oplus\F_{Q^2}(T_{\rm ar})\Pi,
 \qquad \Pi^2=T_{\rm ar},\quad\Pi x=\bar x\Pi.
\]
The algebra ramifies at $0$ and $\infty$ and splits at $\mathfrak p=(T_{\rm ar}-1)$.
At this split place, $PGL_1(\mathcal D)$ acts on the $(Q+1)$-regular Bruhat--Tits tree~\cite[Proposition~2.20]{RSV}.
In a fixed order, let $\alpha$ run through the $D$ norm-one elements of $\F_{Q^2}^{\times}$, parametrized by $(1+\xi\omega)/(1+\xi\bar\omega)$ for $\xi\in\F_Q$, with $\omega/\bar\omega$ at $\xi=\infty$.
The elements $s_\alpha=[1+\alpha\Pi]$ generate a group $\Gamma$ acting simply transitively on the tree vertices~\cite[Proposition~2.21(3) and Corollary~2.23]{RSV}. In this group $s_\alpha^{-1}=s_{-\alpha}$. Its free or free-product presentation is as in~\cite[Theorems~2.24, 2.29]{RSV}. The even-word subgroup is generated projectively by the norm-one principal units
\[
 b_\alpha=\frac{(1-\Pi)(1+\alpha\Pi)}{1-T_{\rm ar}},
 \qquad [b_\alpha]=s_1^{-1}s_\alpha,
\]
since $s_\alpha s_\beta=[b_{-\alpha}]^{-1}[b_\beta]$. The label $\alpha=1$ gives the identity port. These same $D$ labelled units will serve both port directions, even in quotients where distinct labels have equal images.

At the ramified place $0$ put $\mathcal O=\F_{Q^2}[[\Pi]]$, let $U_j$ be the projective image of $1+\Pi^j\mathcal O$, set $\Gamma_j=\Gamma\cap U_j$, and write $\mathcal B=\langle b_\alpha\rangle$. Every $s_\alpha$ has a principal-unit representative. The filtrations are normal, and $U_j/U_{j+1}$ is elementary abelian of characteristic $p$, so the finite quotients are $p$-groups.
For $\Sigma=\{0,\infty,\mathfrak p\}$, let $\Phi_\Sigma=PGL_{1,\mathcal A}(\F_Q[T_{\rm ar},T_{\rm ar}^{-1},(T_{\rm ar}-1)^{-1}])$ be the maximal-order integral group.
By~\cite[Corollary~2.22(3)]{RSV}, the kernel of its action on $\Pi\mathcal O/\Pi^2\mathcal O$ is $\Gamma$.
Every $U_j$ acts trivially on this quotient, giving
\begin{equation}\label{eq:arithmetic-kernel-comparison}
 \Phi_\Sigma\cap U_j=\Gamma\cap U_j=\Gamma_j\quad(j\ge1).
\end{equation}

\subsection{Spectrum and labelled-port transfer}

\begin{lemma}[Ramified-level spectrum]\label{lem:ramified-ramanujan}
For every $n\ge2$, the adjacency operator on $\Gamma_{2n}\backslash\mathcal T_{\mathfrak p}$ has spectrum in $\{D,-D\}\cup[-2\sqrt Q,2\sqrt Q]$.
\end{lemma}
\begin{proof}
We use the integral model to apply the prime-degree Ramanujan bound at a level $I_n$ supported at $0$, with the transfer in positive characteristic. The inclusion $\Phi(I_n)\subseteq\Gamma_{2n}$ then lets us descend the spectrum to $\Gamma_{2n}\backslash\mathcal T_{\mathfrak p}$. The coefficient ring of $\Phi_\Sigma$ inverts $T_{\rm ar}$, so we impose the level over $o_0=\F_Q[\varpi_\infty]$, the ring of functions regular away from $\mathfrak p$, where $\varpi_\infty=(T_{\rm ar}-1)^{-1}$. The order of~\cite[Section~2.2]{RSV} has basis
\[
 \mathcal A(o_0)=o_0\{1,\omega,Z,\omega Z\},\quad
 Z=\varpi_\infty\Pi,\quad
 Z^2=\varpi_\infty(\varpi_\infty+1),\quad Za=\bar aZ.
\]
Conjugation in this basis gives the embedding $\iota:G'=PGL_1(\mathcal D)\hookrightarrow GL_4$ by Skolem--Noether. We use this embedding in~\cite[Sections~3.2 and~5]{LSV}, identifying $G'$ with its image in matrix intersections. With $R_0=o_0$, the arithmetic group and its level-$I$ subgroup in~\cite[Equations~(1.3) and (3.8)]{LSV} are
\[
 \Phi=G'(\mathcal K)\cap GL_4(o_0),\qquad
 \Phi(I)=\{\gamma\in\Phi:\iota(\gamma)\equiv I_4\pmod I\}.
\]
At each place $w\ne\mathfrak p$, the local integer group is $G'(\mathcal K_w)\cap GL_4(\mathcal O_w)$, where $\mathcal O_w$ is the integer ring of the completion $\mathcal K_w$. At $0$ and $\infty$, the coefficient extension is unramified and $Z^2$ has valuation one, so the completed order is the unique maximal order~\cite[proof of Lemma~5.1]{LSV}. Every local conjugation preserves this order, and hence the local integer group equals $G'(\mathcal K_w)$ at these two places.

The $\Sigma$-integer ring is $R_\Sigma=o_0[\varpi_\infty^{-1},(\varpi_\infty+1)^{-1}]$. Since $\varpi_\infty$ is a unit there, the basis above gives $\mathcal A(o_0)\otimes_{o_0}R_\Sigma=\mathcal A(R_\Sigma)$. Each $\gamma\in\Phi$ has an invertible integral conjugation matrix, so it acts by an algebra automorphism of $\mathcal A(o_0)$ and extends to $\mathcal A(R_\Sigma)$. The latter is Azumaya, with automorphism group $PGL_{1,\mathcal A}(R_\Sigma)$ via conjugation~\cite[Sections~2.2--2.3]{RSV}. Thus $\Phi\subseteq\Phi_\Sigma$.

Choose $I_n=(\varpi_\infty+1)^n$, supported at $0$ because $\varpi_\infty+1=\varpi_\infty T_{\rm ar}$. Apply~\cite[Theorem~1.2; Section~5]{LSV} with $k=\mathcal K$, $d=2$, $\nu_0=\mathfrak p$, $R_0=o_0$, and $G'=PGL_1(\mathcal D)$ embedded by $\iota$. The completions at $0,\infty$ are division algebras, so the prime-degree theorem permits every nonzero level $I\lhd o_0$, including $I_n$.

For the transfer in positive characteristic~\cite[Remark~1.6]{LSV}, use~\cite[Theorems~2.5, 3.2]{BR}. Inflate a discrete $PGL_1(\mathcal D)$ representation with infinite-dimensional spherical component at $\mathfrak p$ to $\mathcal D^\times$ with trivial central character. The inverse bijection of~\cite[Theorem~3.2 and Section~3.3]{BR} gives a discrete $GL_2$ representation with the same split-place component. The character identities of~\cite[Theorem~2.5]{BR}, applied also to central multiples, preserve the trivial central character. The infinite-dimensional split-place component rules out the one-dimensional residual representations~\cite[Section~4.1]{LSV}. The transfer is therefore cuspidal and tempered at $\mathfrak p$ by~\cite[Theorem~4.1]{LSV}. Thus~\cite[Proposition~1.5 and Theorem~2.11]{LSV} give adjacency spectrum in $\{D,-D\}\cup[-2\sqrt Q,2\sqrt Q]$ on $\Phi(I_n)\backslash\mathcal T_{\mathfrak p}$.

Let $\gamma\in\Phi(I_n)$. At $0$, $\varpi_\infty$ is a unit, so conjugation by $\gamma$ is the identity modulo $T_{\rm ar}^n$ in the basis $\{1,\omega,\Pi,\omega\Pi\}$. Write $\gamma=\Pi^\epsilon x$ with $\epsilon\in\{0,1\}$ and $x\in\mathcal O^\times$. Modulo $\Pi$, conjugation by a unit is trivial and conjugation by $\Pi$ acts as nontrivial Frobenius on $\F_{Q^2}$. Thus $\epsilon=0$. For $x=a+b\Pi$ with $a,b\in\F_{Q^2}[[T_{\rm ar}]]$, the commutators are
\[
 [x,\omega]=b(\bar\omega-\omega)\Pi,\qquad
 [x,\Pi]=(a-\bar a)\Pi+T_{\rm ar}(b-\bar b).
\]
They lie in $T_{\rm ar}^n\mathcal O$. Since $\bar\omega-\omega$ is a unit, $b\in T_{\rm ar}^n\F_{Q^2}[[T_{\rm ar}]]$ and $a-\bar a\in T_{\rm ar}^n\F_{Q^2}[[T_{\rm ar}]]$. Hence the first $n$ coefficients of $a$ are in $\F_Q$, and $x$ differs modulo $T_{\rm ar}^n$ from a central unit $c\in\F_Q[[T_{\rm ar}]]^\times$. As $T_{\rm ar}^n=\Pi^{2n}$,
\begin{equation}\label{eq:ramified-level-inclusion}
 \Phi(I_n)\subseteq\Phi_\Sigma\cap U_{2n}=\Gamma_{2n}.
\end{equation}
Let $\pi:\Phi(I_n)\backslash\mathcal T_{\mathfrak p}\to\Gamma_{2n}\backslash\mathcal T_{\mathfrak p}$ be the quotient map. By simple transitivity of $\Gamma$, the vertex set for either subgroup $H$ is $H\backslash\Gamma$, with adjacency $(A_Hf)(Hg)=\sum_\alpha f(Hgs_\alpha)$. The map $\pi$ preserves each labelled right translation, so $A_{\Phi(I_n)}\pi^*=\pi^*A_{\Gamma_{2n}}$. Since $\pi$ is onto, $\pi^*$ is injective. Thus the spectrum on $\Gamma_{2n}\backslash\mathcal T_{\mathfrak p}$ is contained in the Ramanujan spectrum on $\Phi(I_n)\backslash\mathcal T_{\mathfrak p}$.
\end{proof}

We now pass from the tree adjacency bound to the labelled averages used by the code construction.

\begin{lemma}[Labelled-port transfer]\label{lem:port-spectral-transfer}
Let $\mathcal B_n$ denote the image of $\mathcal B$ modulo $\Pi^n$ and put $\sigma_D=2\sqrt Q/(Q+1)$. For $n\ge2$ the normalized $D$-labelled left and right port averages on $\mathcal B_n$ have mean-zero operator norms at most $\sigma_D$. The bound descends to every label-preserving quotient, with repeated images counted by their labels.
\end{lemma}
\begin{proof}
In $\bar\Gamma_{2n}=\Gamma/\Gamma_{2n}$, write $(R_\gamma f)(x)=f(x\gamma)$ for right translation and let $P=D^{-1}\sum_\alpha R_{s_\alpha}$ be the normalized tree adjacency. Let $\bar\Gamma_{2n}^+$ be the image of the even-word subgroup of $\Gamma$. The two-step walk $P^2$ preserves this image. Let $T=D^{-1}\sum_\alpha R_{[b_\alpha]}$ there. As $s_\alpha^{-1}=s_{-\alpha}$, $P$ is self-adjoint, and $[b_\alpha]^{-1}[b_\beta]=s_\alpha^{-1}s_\beta$ yields the exact identity
\begin{equation}\label{eq:arithmetic-two-step}
 (T^*Tf)(x)=D^{-2}\sum_{\alpha,\beta}f(xs_\alpha^{-1}s_\beta)
 =(P^2f)(x)\quad(x\in\bar\Gamma_{2n}^+).
\end{equation}
The two-step walk is connected on the even-word image because its support includes each port $s_1^{-1}s_\alpha$. Its $1$-eigenspace therefore consists of constants. The remaining eigenvalues of the restricted $P^2$ come from the nontrivial spectrum in Lemma~\ref{lem:ramified-ramanujan} and are at most $\sigma_D^2$. Therefore $\|T\|_{1^\perp},\|T^*\|_{1^\perp}\le\sigma_D$.

The projective map from $\mathcal B$ to $\Gamma^+$ is injective because a scalar norm-one principal unit has square $1$ and constant coefficient $1$, hence equals $1$.
There is a label-preserving surjection $\bar\Gamma_{2n}^+\twoheadrightarrow\mathcal B_n$.
Indeed, a norm-one principal unit projectively trivial modulo $\Pi^{2n}$ equals $c(1+\Pi^{2n}z)$ for a central $c\in1+T_{\rm ar}\F_Q[[T_{\rm ar}]]$.
The norm-one condition gives $c^2\equiv1\pmod{T_{\rm ar}^n}$.
For odd $p$, $c+1$ is a unit, giving $c\equiv1\pmod{T_{\rm ar}^n}$.
For $p=2$, $c^2-1=(c-1)^2$ gives $c\equiv1\pmod{T_{\rm ar}^{\lceil n/2\rceil}}$.
Since $T_{\rm ar}=\Pi^2$ and $2\lceil n/2\rceil\ge n$, both cases give $c\equiv1\pmod{\Pi^n}$, proving the kernel inclusion.
Also $\mathcal B_1$ is trivial.

For a label-preserving quotient $H\twoheadrightarrow L$, pullback on normalized counting spaces preserves means and norms and intertwines each labelled average and its adjoint. Hence the mean-zero bound descends to $L$. Inversion interchanges left averaging with the adjoint of right averaging. Both operations preserve the labels and their multiplicities.
\end{proof}

\subsection{Effective central tower and numbering}\label{app:effective-central-tower}
\input{sections/appendix-a-code/arithmetic-algorithms}

\input{sections/appendix-a-code/fixed-family-assembly}

%% file: sections/appendix-a-code/arithmetic-algorithms.tex
It remains to construct compatible quotients with central order-$p$ kernels and to make their group data and numbering explicit.
Fix an $\F_p$-basis of $\F_Q$. For a principal unit $x=1+\sum_{j\ge1}c_j\Pi^j$, write $c_j=a_j+b_j\omega$ and order its scalar coordinates by increasing $j$, first those of $a_j$ and then those of $b_j$. Write $c_\nu(x)$ for the $\nu$th scalar coordinate.

\begin{lemma}\label{lem:effective-p-tower}
For every $m\ge0$ there is a compatible label-preserving quotient $G_m$ of $\mathcal B$, of order $p^m$, with central order-$p$ kernels $G_{m+1}\twoheadrightarrow G_m$.
For $m\ge1$, use precision $n=2\lceil m/a\rceil+1$ with this scalar coefficient order. The construction returns the quotient chain, transversals, fibre cycles, labelled port translations, growth positions $\nu_1<\cdots<\nu_m$, and precision-$n$ lifts $\mathfrak z_j$ of the kernel generators. Each $\mathfrak z_j$ has zero earlier scalar coordinates and coordinate $1$ at $\nu_j$.
Writing $[c]_p\in[p]$ for the integer representative of $c\in\F_p$, the growth coordinates define a bijection
\begin{equation}\label{eq:arithmetic-index}
 \operatorname{idx}(x)=\sum_{j=1}^m[c_{\nu_j}(x)]_p p^{j-1}:G_m\longrightarrow[g].
\end{equation}
For $m=0$, set $\operatorname{idx}(1)=0$. For $m\ge1$, reduction modulo $p^{m-1}$ gives the index in $G_{m-1}$.
Given $g=p^m$, all these data and the index of each element are constructible using $T_G=O_D(g(1+\log g))$ arithmetic-field and index operations.
\end{lemma}
\begin{proof}
A norm-one principal unit has expansion $1+\sum_{j\ge1}c_j\Pi^j$. Its truncated image $\mathcal B_n$ has at most $Q^{2(n-1)}$ elements. For the lower bound, let $l=\lfloor(n-1)/2\rfloor$ and take the $Q^l$ products $b_{\beta_1}\cdots b_{\beta_l}$ with norm-one $\beta_i\ne1$. Their corresponding words $s_{-1}s_{\beta_1}\cdots s_{-1}s_{\beta_l}$ are reduced and distinct in the RSV presentation. A common denominator $(1-T_{\rm ar})^l$ and numerator degree at most $2l<n$ show that equality modulo $\Pi^n$ would force equality of numerators and words. Hence
\begin{equation}\label{eq:arithmetic-truncated-size}
 Q^{\lfloor(n-1)/2\rfloor}\le|\mathcal B_n|\le Q^{2(n-1)}.
\end{equation}
Let $\mathcal N_\nu$ consist of the elements of $\mathcal B$ whose first $\nu$ scalar coordinates vanish. Two units $x,y$ have the same prefix exactly when $xy^{-1}\in\mathcal N_\nu$, since multiplication by $y^{-1}$ preserves the first nonzero coefficient of $x-y$. This makes $\mathcal N_\nu$ a subgroup, and conjugation preserves the same first nonzero coefficient, so it is normal. Put $\mathcal P_\nu=\mathcal B/\mathcal N_\nu$. On $\mathcal N_\nu$ the next scalar coordinate is additive and invariant under conjugation, so every nontrivial step $\mathcal P_{\nu+1}\to\mathcal P_\nu$ has central kernel of order $p$. The unbounded sizes in~\eqref{eq:arithmetic-truncated-size} and increments at most $p$ imply that the first prefix attaining $p^m$ exists for every $m$. It defines $G_m$, with the stated compatible kernels.

At precision $n=2\lceil m/a\rceil+1$, the lower bound in~\eqref{eq:arithmetic-truncated-size} gives $|\mathcal B_n|\ge p^m$. The first prefix attaining $p^m$ lies within this truncation, so $G_m$ is a label-preserving quotient of $\mathcal B_n$. For each growing quotient, the algorithm saves representatives at precision $n$ by breadth-first traversal along the labelled ports and their inverses. It computes each generator product once at that precision and compares it with the saved representative of its quotient image. At each prefix before the first disagreement, the representatives map to a set containing the identity and closed under the generators, so they exhaust the quotient. The first differing scalar coordinate splits a class, giving the next central extension. Each group of sizes $1,p,\ldots,p^m$ is enumerated once, and $\sum_{k=0}^mp^k\le2g$.

Each product with a port or its inverse has denominator $1-\Pi^2$ and numerator $q_s(\Pi)$ of degree two. For $b_\alpha$ its coefficients are $(1,\alpha-1,-\bar\alpha)$ and for $b_\alpha^{-1}$ they are $(1,1-\alpha,-\alpha)$. If $x=\sum_jx_j\Pi^j$, then $y=sx$ or $xs$ is obtained from
\[
\begin{aligned}
 y_j-y_{j-2}&=\sum_{t=0}^2(q_s)_t\operatorname{Fr}^{t}(x_{j-t})&&\text{(left)},\\
 y_j-y_{j-2}&=\sum_{t=0}^2x_{j-t}\operatorname{Fr}^{j-t}((q_s)_t)&&\text{(right)},
\end{aligned}
\]
where $\operatorname{Fr}(z)=z^Q$ and negative coefficients vanish. A product and comparison cost $O(an)$ arithmetic-field/index operations. Prefix tries over the fixed alphabet $\F_p$ select representatives and transversals. Ordering a newly split fibre by its first differing $\F_p$ coordinate gives the central $p$-cycle. In the fibre over the identity, the representative with new coordinate $1$ supplies $\mathfrak z_j$. Each step has exactly $p$ extensions, so the $m$ growth coordinates range bijectively over $\F_p^m$, proving~\eqref{eq:arithmetic-index}. Reading these coordinates for all enumerated representatives takes $O(gm)$ index operations, included in $T_G$. The labelled left/right translations and inverses are tabulated at each quotient. There are $O(Dg)$ retained entries over the chain and $O(agn)$ coefficient/index cells for representatives, giving $O(aDgn)=O_D(g(1+\log g))$ work. Appendix~\ref{app:addressing} specifies public compact addressing and recovery-side permutation-table construction from this interface, including their costs.
\end{proof}

%% file: sections/appendix-a-code/fixed-family-assembly.tex
\subsection{Assembly of the tower and code family}\label{app:family-assembly}
\begin{proof}[Lemma~\ref{lem:arithmetic-spectrum}]
\phantomsection\label{proof:arithmetic-spectrum}
Lemma~\ref{lem:effective-p-tower} supplies the central quotient data, public indices, and preparation bound $T_G$.
Lemma~\ref{lem:ramified-ramanujan} establishes the level spectrum. For $m\ge1$, Lemma~\ref{lem:port-spectral-transfer} transfers the norm bound to $G_m$ through the truncation identified in Lemma~\ref{lem:effective-p-tower}. Thus $\sigma_D=2\sqrt{D-1}/D$ bounds the mean-zero operator norms in both directions, including on the trivial group $G_0$.
\end{proof}

\begin{proof}[Proposition~\ref{prop:family}]
Fix $\bar r$ and $\bar e$ as in the \hyperref[app:target-rate-choice]{target-rate proof in Appendix~\ref*{app:fixed-field-checks}} and choose a sufficiently large $D=p^a+1$. Let $r,e$ and $d=d_{\cap}$ be as in Lemma~\ref{lem:qualified-local-checks}, which supplies fixed checks meeting the typed and intersection distances and $J_e$. As $D\to\infty$, $d/D\to4\bar e>2\bar e$, $e/D\to\bar e>0$, and $\sigma_D=2\sqrt{D-1}/D\to0$, so $e\ge2$, $d\ge2e-1$, $d/D>\sigma_D$, and $e+1>2D\sigma_D$ hold for sufficiently large $D$. The last inequality gives the strict margin for shared completion. All $G_m$ inherit the same port bounds by Lemma~\ref{lem:arithmetic-spectrum}.

Since $d=d_{\cap}$, we may set $d_A^{\cap}=d_B^{\cap}=d$ as well as $d_A=d_B=d$, and Proposition~\ref{prop:distance} yields the common positive relative distance
\[
 \Delta=\frac{d}{D}\,
 \frac{(d/D)(d/D-\sigma_D)}{1-\sigma_D}>0.
\]
Theorem~\ref{thm:encoding} gives $N=D^2p^m$ and $K=(D-2r)^2p^m$, and its whole-kernel encoder together with the tower preparation uses $O_{\rm fam}(N\log N)$ operations, counting the two fields separately. The constants in Theorem~\ref{thm:testing} are independent of $m$, as is the positive normalized closure capacity $\Lambda_e/(4p^m)=((e+1)/(2D)-\sigma_D)/(1-\sigma_D)$. The rate bound in Lemma~\ref{lem:qualified-local-checks} gives $K/N=(1-2r/D)^2\ge\rho$. The two fields, $D$, port list, local checks, and qualification bounds remain fixed along the tower.
\end{proof}

\subsection{Verification of the finite parameters}\label{app:finite-recovery-parameters}
For the local parameters of Section~\ref{sec:finite-instance}, using $B_q(D,w)\le (w+1)\binom Dw q^w$ for $w\le D/2$ gives $\epsilon_D<2^{-3190}$ in~\eqref{eq:fullrank-finite-setup}. Thus Lemma~\ref{lem:finite-local-qualification} supplies checks with the stated distances and $J_e$. Lemma~\ref{lem:arithmetic-spectrum} gives $g=2$ and $D\sigma_X\le182$ for both axes. Since $e+1>2D\sigma_{\rm avg}$, the instance is qualified. At $N=134{,}250{,}498$, Proposition~\ref{prop:distance} gives $d(\C)\ge3{,}178{,}225$, so we take $d_C=3{,}178{,}225$.

Equation~\eqref{eq:recovery-hole-budgets} gives $L_*=q_*=n_{\rm hole}=0$. Then~\eqref{eq:recovery-residual} gives $\widehat{\mathsf T}=\lfloor105^2/16\rfloor=689$. Admissibility follows from $2(105-1)+1690=1898<1900$, $2n_{\rm hole}=0<\Lambda_e$, and $2(t+u)=2{,}801{,}378<d_C$, so Corollary~\ref{cor:instance-interface} applies.

With $\Upsilon=133\ln2$, Equation~\eqref{eq:closed-coverage-allocation} gives $n_V=100{,}000$. At this draw count, the first bound in~\eqref{eq:closed-heavy-allocation} and the avoidance bound~\eqref{eq:lt-prob} also give errors at most $2^{-133}$ each. With $d_{\rm tree}=28$, the base query count in Appendix~\ref{app:full-resource-ledger} is at most $49{,}426{,}500{,}996$. The cap and digest width satisfy~\eqref{eq:fixed-das-digest-width} at $\lambda=128$. Theorem~\ref{thm:das} then gives the claimed security.

%% file: sections/appendix-a-code/client-addressing.tex
\section{Public Addressing and Recovery Cost}\label{app:addressing}\label{app:client-addressing}\label{app:recovery-addressing}

The arithmetic construction supplies the index and group data; the following result gives the additional addressing data and their costs.

\begin{lemma}\label{lem:public-addressing}
Given the data of Lemma~\ref{lem:effective-p-tower} for fixed $Q=p^a=D-1$, with $g=p^m$ and precision $n=2\lceil m/a\rceil+1$ (so $n=1$ when $m=0$), the following bounds hold.
\begin{enumerate}
\item \emph{Client addressing.} Compact client data take $O_D((1+m)^2)$ bits. Preparing the saved generator powers takes $O(pmn^2)$ arithmetic-field operations. Each uniform line draw can be converted into its ordered coordinate list in $O_D(mn^2+D(n+m))=O_{\rm fam}((1+\log N)^3)$ arithmetic-field and index operations.
\item \emph{Recovery tables.} Tables for $x\mapsto a_i^{\pm1}x$ and $x\mapsto xb_j^{\pm1}$ can be compiled in an additional $O_D(Dgn)$ arithmetic-field and index operations. They store $4Dg$ indices, or $4Dg\lceil\log_2g\rceil$ bits for $g\ge2$, and are implicit for $g=1$.
\item \emph{Publication order.} Arranging each publication's information and output symbols in the public orders takes $O_D(N(1+m))$ index operations.
\end{enumerate}
The compact data and recovery tables can be reused across publications; the reordering cost is charged to each publication.
\end{lemma}

\begin{proof}
\textbf{Client addressing.}
Using the lifts from Lemma~\ref{lem:effective-p-tower}, store their powers $\mathfrak z_j^t$ for $0\le t<p$. Given base-$p$ digits $\mathsf d_j\in[p]$, recover the representative by starting at $x=1$ and applying the updates in order:
\[
 x\leftarrow x\mathfrak z_j^{(\mathsf d_j-[c_{\nu_j}(x)]_p)\bmod p}\pmod{\Pi^n}
 \qquad(j=1,\ldots,m).
\]
Multiplying by $\mathfrak z_j^t$ adds $t$ modulo $p$ to $[c_{\nu_j}(x)]_p$ and fixes earlier scalar coordinates. After step $j$ the first $j$ digits match, so the loop inverts $\operatorname{idx}$. For $m\ge1$, the powers, growth positions, and fixed field/port data use $O(pmn)$ $\F_{Q^2}$ cells, or $O_D((1+m)^2)$ bits for fixed $Q$. Computing the powers from saved lifts costs $O(pmn^2)$ arithmetic-field operations. A client generating these data locally from scratch first constructs the tower at cost $T_G$. Decoding uses $m$ truncated products at $O(n^2)$ operations each, for $O(mn^2)$ total.

For a uniform line draw, choose its direction, type, port, and group index uniformly and independently. Decode the group index once. Each of the $D$ labelled translations then costs $O_D(n)$ operations by the degree-two product recurrences in Appendix~\ref{app:effective-central-tower}. Reading the $m$ growth coordinates of the translated representative gives its index by~\eqref{eq:arithmetic-index} in $O(m)$ index operations, and~\eqref{eq:public-leaf-order} gives its coordinate address. This gives the ordered coordinate lists of Section~\ref{sec:source-hypotheses}, with separate labels for repeated port images. Per draw it uses
\[
 O_D(mn^2+D(n+m))=O_{\rm fam}((1+\log N)^3)
\]
arithmetic-field and index operations. The $m=0$ case is immediate.

\textbf{Recovery tables.}
The recovery party materializes
\[
 P_A^{\pm}(i,\operatorname{idx}(x))=\operatorname{idx}(a_i^{\pm1}x),\qquad
 P_B^{\pm}(j,\operatorname{idx}(x))=\operatorname{idx}(xb_j^{\pm1}).
\]
Starting from the representatives saved during tower preparation, the same recurrences and growth-coordinate reads cost $O_D(n+m)=O_D(n)$ per forward image at the full quotient. There are $2Dg$ forward images across the two directions, for a total of $O_D(Dgn)$ arithmetic-field and index operations. The maps are permutations at each label, so set $P_X^-(\tau,P_X^+(\tau,k))=k$. Each line takes at most $D$ lookups. The four tables hold $4Dg$ indices or $4Dg\lceil\log_2g\rceil$ packed bits for $g\ge2$, and are implicit for $g=1$.

\textbf{Publication order.}
For publication, using the prepared indices to arrange the information and output symbols in~\eqref{eq:public-message-order} and~\eqref{eq:public-leaf-order} costs $O_D(N(1+m))$ index operations.
\end{proof}

\input{sections/appendix-b-recovery/recovery-implementation}

%% file: sections/appendix-b-recovery/recovery-implementation.tex
\subsection{Linear work and state}\label{app:recovery-implementation}
With the translation tables constructed above, we can bound the work and workspace of checked recovery.

\begin{proposition}[Recovery cost]\label{prop:recovery-cost}
Fix a qualified instance with admissible parameters and $r_{\max}=\max(r_A,r_B)$. Suppose the public indices have tables for $x\mapsto a_i^{\pm1}x$ and $x\mapsto xb_j^{\pm1}$. Then recovery uses $O(DN+r_{\max}h(N-|S|))$ code-field operations and $O(N)$ table lookups/index operations on $y\in\F^S$. Dynamic storage is $O(N)$ field cells and index words plus $O(D^2)$ reusable temporary field cells, with $O(\log N)$ bits per index. Thus post-preparation algebraic work is $O_{\rm fam}(N)$, excluding transcripts and authentication.
\end{proposition}
\begin{proof}
Here $N=D^2g$, $MD=4N$, and $r_{\max}<D/2$ by the construction. The compiled permutations give each line/view incidence in at most two lookups, for $4N$ entry addresses and $O(Dg)$ labelled-neighbour updates. On a non-heavy line with $u_{\mathcal L}\le h$ missing symbols, computing the syndrome and solving~\eqref{eq:partial-line} cost $O(r_{\max}D+r_{\max}u_{\mathcal L}^2)$ code-field operations. The inequality $h<d_{\min}$ makes any solution unique. Each missing physical coordinate occurs on four lines, so
\[
 \sum_{\mathcal L}u_{\mathcal L}=4(N-|S|),\qquad
 \sum_{u_{\mathcal L}\le h}u_{\mathcal L}^2\le4h(N-|S|).
\]
This gives the claimed line cost. A checked tensor lift tests the supplied rectangle, selects full-rank retained minors, solves the two component systems, and verifies the output in $O(D^3)$ field operations using $O(D^2)$ scratch cells. Preparing the generator matrices of the four local codes costs $O(D^3)$ field operations per instance.

The first pass records the initial holes using temporary lifts. After closure, the second pass materializes the surviving views and fills the holes. The procedure initializes first-in, first-out queues in fixed vertex order and scans threshold updates in fixed port order. Each view is lifted at most twice and enters each applicable queue at most once. Each labelled incidence is updated $O(1)$ times. The $4g$ views therefore cost $O(D^3g)=O(DN)$ field work and $O(N)$ index work.

Merging the four corners per coordinate and checking the line equations, erasure bound, and observed disagreement add $O(DN)$ field and $O(N)$ index operations. Completion masks, stored lines and views, queues, counters, and output fit in $O(N)$ dynamic cells and words. The resource ledger also charges compiled tables, group construction, and authentication.
\end{proof}

%% file: sections/appendix-b-recovery/collective-das-supplement.tex
\section{Fixed-Family Budgets, Sampling, and Resources}\label{app:collective-das}
For Section~\ref{sec:security}, we verify the fixed-family budgets, derive an explicit sampling threshold, and give the complete resource bounds.

\subsection{Admissibility of the fixed-family budgets}\label{app:fixed-budget-check}
Use the margins~\eqref{eq:family-admissible-fractions} and the integer budgets~\eqref{eq:family-positive-budgets}. The choice in Section~\ref{sec:family-guarantee} gives $2n_{\rm hole}<\Lambda_e$; it remains to check the residual and distance inequalities of~\eqref{eq:admissible-basic} and the stated orders of the budgets.

At $(H_0,I_0)=(2L_*,2q_*)$, use $\mathsf C_e(s)<2s$ for $s>0$ and $\mathsf C_e(0)=0$ to get at most $D^2\mathsf C_e(2L_*+2q_*)\le4D^2(L_*+q_*)$ corners inside closure.
The second entry of the minimum in~\eqref{eq:recovery-corner-bound} bounds the outside count by $\kinit[\ill_{\max}-(\kinit+1)q_*]/2$.
Local consistency gives $\kinit\le d_{\min}\le D$, hence $\kinit/8\le D^2/(\kinit+1)$.
\vadjust pre{\hbox{\phantomsection\label{app:fixed-residual-estimate}}}%
Dividing by four and applying this inequality gives
\[
\begin{aligned}
 \widehat{\mathsf T}
 &\le D^2(L_*+q_*)+\frac{\kinit}{8}\bigl[\ill_{\max}-(\kinit+1)q_*\bigr]\\
 &\le D^2\left(L_*+\frac{\ill_{\max}}{\kinit+1}\right)
 \le N(C_{\rm res}\theta+\eta)\le t.
\end{aligned}
\]
Put $\nu=\Delta-2(C_{\rm res}\theta+\eta+\zeta)>0$.
For $g>\max\{2/(D^2\nu),2/(D^2\zeta),1/(2D\theta),2/\eta\}$, each floor in~\eqref{eq:family-positive-budgets} retains at least half of its argument.
Also $N\nu>2$, so $2(t+u)\le N(\Delta-\nu)+2<d_C=\lceil N\Delta\rceil$.

\input{sections/appendix-b-recovery/occupancy}

\subsection{An explicit draw threshold}\label{app:asymptotic-details}
Combining the observation tails of Appendix~\ref{app:occupancy} with the illegal-line avoidance bound~\eqref{eq:lt-prob} gives one explicit sufficient choice of $n_{\rm req}$ for Lemma~\ref{lem:family-budgets}.
\begin{proof}[Explicit draw threshold]

Use the fixed fractions $\theta,\eta$ of~\eqref{eq:family-admissible-fractions} and budgets~\eqref{eq:family-positive-budgets}. Section~\ref{sec:family-guarantee} and Appendix~\ref{app:fixed-budget-check} establish admissibility above a member-size threshold independent of $\lambda$. Let $N=D^2g$ and $M=4Dg$.

Given $\lambda\ge1$, put $\Upsilon=(\lambda+4)\ln2$ and take $n_V$ from~\eqref{eq:closed-coverage-allocation}. For $R<M$, the second bound in~\eqref{eq:closed-heavy-allocation} is at most $e^{-\Upsilon}$ at
\[
 n_H=\left\lceil\frac{M}{R+1}\bigl(M\ln2+\Upsilon\bigr)\right\rceil .
\]
Take $n_H=0$ if $R\ge M$. A sufficient threshold is
\begin{equation}\label{eq:family-line-only-counts}
 n_{\rm req}=\max\!\left\{\left\lceil\Upsilon/[-\ln(1-\theta)]\right\rceil,\ n_V,\ n_H\right\}.
\end{equation}
For a registered $W$ with $\ill(W)>\ill_{\max}$, each draw avoids illegal lines with probability at most $1-\theta$, so the first term bounds the avoidance error by $e^{-\Upsilon}$. The other two terms give observation error at most $2e^{-\Upsilon}=2^{-(\lambda+3)}$. Since $u+1>\zeta N$, using $\delta=1+a$ in~\eqref{eq:closed-coverage-allocation} gives $n_V=O_{\rm fam,\zeta}(N/D+\Upsilon)$. Since $R+1>\eta g$, we have $M/(R+1)<4D/\eta$, and $M=4N/D$ gives the same order for $n_H$. Thus $n_{\rm req}\le c_{\rm draw}(N/D+\lambda)$, with recovery budgets independent of $\lambda$.
\end{proof}

\input{sections/appendix-b-recovery/full-resource-ledger}

%% file: sections/appendix-b-recovery/occupancy.tex
\subsection{Coverage and heavy-line tails}\label{app:occupancy}
Condition on the registration history and the selected client set. Let $V=N-|S^*|$ and $H=r_{\rm H}(S^*)$, with $0\le h<D$. Write $Z_\beta$ for the indicator that line label $\beta$ is missed, and put $\pi=\exp(-n_{\rm dr}/M)$. For any $b$ distinct labels the probability of missing all of them is $(1-b/M)^{n_{\rm dr}}\le \pi^b$. Since $Z_\beta^2=Z_\beta$, comparing monomials shows that a polynomial in the missed-label indicators with nonnegative coefficients has expectation at most its expectation under independent Bernoulli indicators of mean $\pi$.

For $z\ge1$, expand $z^{V/D}$ as a product of factors $1+(z^{1/D}-1)\prod_{\beta\ni x}Z_\beta$, one for each coordinate $x$. Under the independent comparison an erased coordinate requires four missed labels, and each label enters exactly $D$ factors. Finner's inequality~\cite{Finner} gives
\begin{equation}\label{eq:observation-mgf}
 \mathbb E z^{V/D}\le (1-\pi^4+\pi^4z)^{N/D}.
\end{equation}
Set $s=u+1\le N$. If $\pi^4\le (s/N)e^{-\delta}$ for $\delta\ge0$, Markov's inequality with $z=e^\delta$ and $\ln(1+x)\le x$ give
\[
 \Pr[V\ge s]\le\exp[-(s/D)(\delta-1+e^{-\delta})].
\]
We have $\delta-1+e^{-\delta}\ge\max\{\delta-1,\delta^2/[2(1+\delta)]\}$, with the second bound obtained by integrating $1-e^{-x}\ge x/(1+\delta)$ on $[0,\delta]$. Either $\delta=1+a$ or $\delta=a+\sqrt{a^2+2a}$ makes the probability bound at most $e^{-\Upsilon}$ for $a=D\Upsilon/s$. Since $M=4N/D$ gives $\pi^4=e^{-n_{\rm dr}D/N}$, the condition on $\pi^4$ holds for $n_{\rm dr}\ge n_V$, where
\begin{equation}\label{eq:closed-coverage-allocation}
 n_V=\left\lceil\frac ND\left(\ln(N/s)+\min\{1+a,\ a+\sqrt{a^2+2a}\}\right)\right\rceil.
\end{equation}

Fix $k=h+1$ positions on a line. If all are erased, the draws miss the line's label, the parallel label of the other type through the first chosen position, and $2k$ crossing labels. These crossing labels differ by port across positions and by type at each position, even when port images coincide. Thus a heavy line witnesses at least $2h+4$ missed labels. Also, any $R+1$ heavy lines have $R+1$ distinct missed labels. Union bounds over the corresponding choices give, for $0\le R<M$,
\begin{equation}\label{eq:closed-heavy-allocation}
 \Pr[H>R]\le
 \min\left\{M\binom{D}{h+1}\pi^{2h+4},\ \binom{M}{R+1}\pi^{R+1}\right\}.
\end{equation}
For $R\ge M$ the event is empty. If $n_H$ draws make either bound at most $e^{-\Upsilon}$, then $\max\{n_V,n_H\}$ draws make the observation error in~\eqref{eq:observation-failure} at most $2e^{-\Upsilon}$.

%% file: sections/appendix-b-recovery/full-resource-ledger.tex
\subsection{Complete resource bounds}\label{app:full-resource-ledger}
For a list of $\ell$ clients with complete canonical replies, let $n_{\rm dr}=a_{\rm cl}\ell$, $w=\lceil\log_2q\rceil$, $w_{\rm lab}=\lceil\log_2M\rceil$, and tree depth $d_{\rm tree}=\lceil\log_2N\rceil$.
Client $i$'s draws specify $s_i^{\rm raw}\le a_{\rm cl}D$ coordinate occurrences.
The distinct positions form $S_i^*$.
The reply contains $s_i=|S_i^*|$ values and $n_i^{\rm help}=n_{\rm help}(S_i^*)$ helper digests.
Put $n_{\rm pay}=\sum_i s_i$.

\paragraph{Communication and public data.}
The first hop carries draw labels to the publisher and values and helper digests to the clients. Forwarding the transcripts repeats these items on the second hop. Including context and commitment delivery, these transmissions are bounded by
\begin{equation}\label{eq:combined-communication}
 2\left[n_{\rm dr}w_{\rm lab}+\sum_i(s_iw+\kappa n_i^{\rm help})\right]
 +(\ell+1)(|\operatorname{enc}(\mathsf{ctx})|+\kappa).
\end{equation}
\phantomsection\label{app:public-data-charges}
Per recipient, the fixed checks occupy $O(D^2w)$ bits and the compact addressing data occupy $O_{\rm fam}((1+\log N)^2)$ bits (Appendix~\ref{app:client-addressing}). We charge delivery of the checks and either delivery or local generation of the addressing data. Parties can reuse these data across publications.

\paragraph{Preparation, publication, and storage.}
For the arithmetic family, write $Q=p^a$, $g=p^m$, and $n=2\lceil m/a\rceil+1$. Reusable preparation comprises $T_G$ for the tower, quotient-chain data, and canonical indices (Lemma~\ref{lem:arithmetic-spectrum}), $O(pmn^2)$ arithmetic-field operations for compact client data, and $O_D(Dgn)$ arithmetic-field and index operations for the four recovery tables (Lemma~\ref{lem:public-addressing}). Together these use $O_{\rm fam}(N\log N)$ arithmetic-field and index operations. Appendices~\ref{app:effective-central-tower} and~\ref{app:addressing} give the tower and addressing workspace bounds. The four arrays store $4Dg$ group indices, or $O(Dg\log(g+1))$ bits. Fixed qualified checks incur a once-per-family qualification cost depending on $(D,q)$, as in Appendix~\ref{app:fullrank-asymptotic}. Each publication separately incurs $O_{\rm fam}(N\log N)$ code-field and index encoding work, $O_D(N(1+m))$ index operations for reordering, and exactly $2N$ Merkle hashes including the outer root. Tree storage is $(2N-1)\kappa$ digest bits in addition to $Nw$ codeword bits.

\paragraph{Clients and extraction.}
For $a_{\rm cl}$ draws, on-demand addresses cost $O_D(a_{\rm cl}((1+\log_pg)^3+D(1+\log_pg)))$ arithmetic-field and index operations. Line checks take at most $a_{\rm cl}r_{\max}D$ code-field multiplications and $a_{\rm cl}r_{\max}(D-1)$ additions. Forming $S_i^*$ takes $O(s_i^{\rm raw}\log(s_i^{\rm raw}+1))$ comparisons. Proof construction or verification visits $O(s_i+n_i^{\rm help})$ nodes. With compiled tables, recovery-side line/view addressing uses $O(N+n_{\rm dr}D)$ lookups, merging $O(N+n_{\rm pay})$ code-field and index operations, and checked algebraic recovery $O(DN+r_{\max}hU_{\rm miss})$ code-field operations for $U_{\rm miss}=N-|S|$ (Proposition~\ref{prop:recovery-cost}). Together these three tasks use $O_{\rm fam}(N+n_{\rm dr})$ code-field and index operations. Parsing and serialization count their encoded lengths.

\paragraph{Oracle calls and bit work.}
By~\eqref{eq:multiproof-identity} in Appendix~\ref{app:multiproof-sizes}, verifying a client reply takes $2s_i+n_i^{\rm help}$ hashes at the client and again at extraction. Adding the hashes for one publication gives the base charge for one complete list from the selected clients
\begin{equation}\label{eq:all-authentication-hashes}
 H_{\rm base}=2N+2\sum_i(2s_i+n_i^{\rm help}).
\end{equation}
For one-line clients, $n_i^{\rm help}\le Dd_{\rm tree}$ by~\eqref{eq:multiproof-frontier}, so $H_{\rm base}\le2N+2\ell D(d_{\rm tree}+2)$ independently of $\kappa$.
For $\ell=n_{\rm req}$, this upper bound covers honest publication and both verification passes. Larger lists are supported only if they meet the same public resource bounds.

Fix $\Qhash$ independently of $\kappa$ to bound all queries in the complete experiment. Every call counts toward the cap, whether it comes from a complete honest run, a failed or repeated attempt, or the adversary. Equation~\eqref{eq:fixed-das-digest-width} gives a sufficient digest width for this cap. Bit work counts the encoded hash inputs, including leaf values and labels, typed headers, child digests, context, and root data.

\subsection{Opening-cost comparison}\label{app:cross-scheme-comparison}
We specialize the preceding counts to the publisher--client hop in Table~\ref{tab:sampling-costs}. Replies to $n_{\rm dr}$ draws carry at most $Dn_{\rm dr}$ symbols and $Dn_{\rm dr}\lceil\log_2N\rceil$ helper digests by~\eqref{eq:multiproof-frontier}. Including draw labels gives $O(D\log q+D\kappa\log N)$ bits per draw.
For $n_{\rm req}$ one-line clients, our first hop uses $O_{\rm fam,\zeta}((N/D+\lambda)\log N)$ bits for labels and values and $O_{\rm fam,\zeta}((N/D+\lambda)\kappa\log N)$ for authentication.
Since honest publication takes $2N$ queries, $\log\Qhash=O(\lambda)$ implies $\log K=O(\lambda)$ for $K=\Theta_{\rm fam}(N)$. At the digest threshold in~\eqref{eq:fixed-das-digest-width}, the first-hop sampling exchange is $O_{\rm fam,\zeta}((K+\lambda)\lambda\log K)$ bits.

The $N/D$ term is necessary for coverage at fixed $\zeta$: covering $N-u$ symbols for $u=\lfloor\zeta N\rfloor$ requires $n_{\rm dr}\ge(N-u)/D\ge(1-\zeta)N/D$.